\documentclass[longauth]{aa} 

\usepackage{graphicx}
\usepackage{txfonts}
\usepackage{tabularx}
\usepackage{multirow}
\usepackage{booktabs}
\usepackage{float}
\usepackage{xcolor}
\usepackage{natbib}
\bibpunct{(}{)}{;}{a}{}{,} 
\usepackage[breaklinks, colorlinks, citecolor=blue, linkcolor=blue]{hyperref}
\makeatletter
\renewcommand*\aa@pageof{, page \thepage{} of \pageref*{LastPage}}
\makeatother

\newlength{\bibitemsep}
\newlength{\bibparskip}
\let\oldthebibliography\thebibliography
\renewcommand\thebibliography[1]{%
  \oldthebibliography{#1}%
  \setlength{\parskip}{\bibitemsep}%
  \setlength{\itemsep}{\bibparskip}%
}

\begin{document}

\title{The geometric albedo of the hot Jupiter HD\,189733b measured with CHEOPS \thanks{Based on data from CHEOPS Guaranteed Time Observations, collected under Programme IDs CH\textunderscore PR100016 and CH\textunderscore PR100019. The raw and detrended photometric time-series data are available in electronic form
at the CDS via anonymous ftp to cdsarc.cds.unistra.fr (130.79.128.5)
or via \url{https://cdsarc.cds.unistra.fr/cgi-bin/qcat?J/A+A/}}}

\subtitle{}

\author{A. F. Krenn\inst{1,2} \thanks{andreas.krenn@oeaw.ac.at} $^{\href{https://orcid.org/0000-0003-3615-4725}{\includegraphics[scale=0.5]{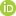}}}$
\and
M. Lendl\inst{2} $^{\href{https://orcid.org/0000-0001-9699-1459}{\includegraphics[scale=0.5]{figures/orcid.jpg}}}$ \and 
J. A. Patel\inst{3} $^{\href{https://orcid.org/0000-0001-5644-6624}{\includegraphics[scale=0.5]{figures/orcid.jpg}}}$ \and 
L. Carone\inst{1} $^{\href{https://orcid.org/0000-0001-9355-3752}{\includegraphics[scale=0.5]{figures/orcid.jpg}}}$ \and 
M. Deleuil\inst{4} $^{\href{https://orcid.org/0000-0001-6036-0225}{\includegraphics[scale=0.5]{figures/orcid.jpg}}}$ \and 
S. Sulis\inst{4} $^{\href{https://orcid.org/0000-0001-8783-526X}{\includegraphics[scale=0.5]{figures/orcid.jpg}}}$ \and 
A. Collier Cameron\inst{5} $^{\href{https://orcid.org/0000-0002-8863-7828}{\includegraphics[scale=0.5]{figures/orcid.jpg}}}$ \and 
A. Deline\inst{2} \and 
P. Guterman\inst{4,6} \and 
D. Queloz\inst{7,8} $^{\href{https://orcid.org/0000-0002-3012-0316}{\includegraphics[scale=0.5]{figures/orcid.jpg}}}$ \and 
L. Fossati\inst{1} $^{\href{https://orcid.org/0000-0003-4426-9530}{\includegraphics[scale=0.5]{figures/orcid.jpg}}}$ \and 
A. Brandeker\inst{3} $^{\href{https://orcid.org/0000-0002-7201-7536}{\includegraphics[scale=0.5]{figures/orcid.jpg}}}$ \and 
K. Heng\inst{9,10,11} $^{\href{https://orcid.org/0000-0003-1907-5910}{\includegraphics[scale=0.5]{figures/orcid.jpg}}}$ \and 
B. Akinsanmi\inst{2} $^{\href{https://orcid.org/0000-0001-6519-1598}{\includegraphics[scale=0.5]{figures/orcid.jpg}}}$ \and 
V. Adibekyan\inst{12,13} $^{\href{https://orcid.org/0000-0002-0601-6199}{\includegraphics[scale=0.5]{figures/orcid.jpg}}}$ \and 
A. Bonfanti\inst{1} $^{\href{https://orcid.org/0000-0002-1916-5935}{\includegraphics[scale=0.5]{figures/orcid.jpg}}}$ \and 
O. D. S. Demangeon\inst{12,13} $^{\href{https://orcid.org/0000-0001-7918-0355}{\includegraphics[scale=0.5]{figures/orcid.jpg}}}$ \and 
D. Kitzmann\inst{14} $^{\href{https://orcid.org/0000-0003-4269-3311}{\includegraphics[scale=0.5]{figures/orcid.jpg}}}$ \and 
S. Salmon\inst{2} $^{\href{https://orcid.org/0000-0002-1714-3513}{\includegraphics[scale=0.5]{figures/orcid.jpg}}}$ \and 
S. G. Sousa\inst{12} $^{\href{https://orcid.org/0000-0001-9047-2965}{\includegraphics[scale=0.5]{figures/orcid.jpg}}}$ \and 
T. G. Wilson\inst{5} $^{\href{https://orcid.org/0000-0001-8749-1962}{\includegraphics[scale=0.5]{figures/orcid.jpg}}}$ \and 
Y. Alibert\inst{15} $^{\href{https://orcid.org/0000-0002-4644-8818}{\includegraphics[scale=0.5]{figures/orcid.jpg}}}$ \and 
R. Alonso\inst{16,17} $^{\href{https://orcid.org/0000-0001-8462-8126}{\includegraphics[scale=0.5]{figures/orcid.jpg}}}$ \and 
G. Anglada\inst{18,19} $^{\href{https://orcid.org/0000-0002-3645-5977}{\includegraphics[scale=0.5]{figures/orcid.jpg}}}$ \and 
T. Bárczy\inst{20} $^{\href{https://orcid.org/0000-0002-7822-4413}{\includegraphics[scale=0.5]{figures/orcid.jpg}}}$ \and 
D. Barrado Navascues\inst{21} $^{\href{https://orcid.org/0000-0002-5971-9242}{\includegraphics[scale=0.5]{figures/orcid.jpg}}}$ \and 
S. C. C. Barros\inst{12,13} $^{\href{https://orcid.org/0000-0003-2434-3625}{\includegraphics[scale=0.5]{figures/orcid.jpg}}}$ \and 
W. Baumjohann\inst{1} $^{\href{https://orcid.org/0000-0001-6271-0110}{\includegraphics[scale=0.5]{figures/orcid.jpg}}}$ \and 
M. Beck\inst{2} $^{\href{https://orcid.org/0000-0003-3926-0275}{\includegraphics[scale=0.5]{figures/orcid.jpg}}}$ \and 
T. Beck\inst{15} \and
W. Benz\inst{14,15} $^{\href{https://orcid.org/0000-0001-7896-6479}{\includegraphics[scale=0.5]{figures/orcid.jpg}}}$ \and 
N. Billot\inst{2} $^{\href{https://orcid.org/0000-0003-3429-3836}{\includegraphics[scale=0.5]{figures/orcid.jpg}}}$ \and 
L. Blecha\inst{22} \and
X. Bonfils\inst{23} $^{\href{https://orcid.org/0000-0001-9003-8894}{\includegraphics[scale=0.5]{figures/orcid.jpg}}}$ \and 
L. Borsato\inst{24} $^{\href{https://orcid.org/0000-0003-0066-9268}{\includegraphics[scale=0.5]{figures/orcid.jpg}}}$ \and 
C. Broeg\inst{15,25} $^{\href{https://orcid.org/0000-0001-5132-2614}{\includegraphics[scale=0.5]{figures/orcid.jpg}}}$ \and 
J. Cabrera\inst{26} $^{\href{https://orcid.org/0000-0001-6653-5487}{\includegraphics[scale=0.5]{figures/orcid.jpg}}}$ \and 
S. Charnoz\inst{27} $^{\href{https://orcid.org/0000-0002-7442-491X}{\includegraphics[scale=0.5]{figures/orcid.jpg}}}$ \and 
C. Corral van Damme\inst{28} \and
Sz. Csizmadia\inst{26} $^{\href{https://orcid.org/0000-0001-6803-9698}{\includegraphics[scale=0.5]{figures/orcid.jpg}}}$ \and 
P. E. Cubillos\inst{1,29} $^{\href{https://orcid.org/0000-0002-1347-2600}{\includegraphics[scale=0.5]{figures/orcid.jpg}}}$ \and 
M. B. Davies\inst{30} $^{\href{https://orcid.org/0000-0001-6080-1190}{\includegraphics[scale=0.5]{figures/orcid.jpg}}}$ \and 
L. Delrez\inst{31,32} $^{\href{https://orcid.org/0000-0001-6108-4808}{\includegraphics[scale=0.5]{figures/orcid.jpg}}}$ \and 
B.-O. Demory\inst{14,15} $^{\href{https://orcid.org/0000-0002-9355-5165}{\includegraphics[scale=0.5]{figures/orcid.jpg}}}$ \and 
D. Ehrenreich\inst{2,33} $^{\href{https://orcid.org/0000-0001-9704-5405}{\includegraphics[scale=0.5]{figures/orcid.jpg}}}$ \and 
A. Erikson\inst{26} \and
J. Farinato\inst{24} $^{\href{https://orcid.org/0000-0002-5840-8362}{\includegraphics[scale=0.5]{figures/orcid.jpg}}}$ \and 
A. Fortier\inst{14,15} $^{\href{https://orcid.org/0000-0001-8450-3374}{\includegraphics[scale=0.5]{figures/orcid.jpg}}}$ \and 
M. Fridlund\inst{34,35} $^{\href{https://orcid.org/0000-0002-0855-8426}{\includegraphics[scale=0.5]{figures/orcid.jpg}}}$ \and 
D. Gandolfi\inst{36} $^{\href{https://orcid.org/0000-0001-8627-9628}{\includegraphics[scale=0.5]{figures/orcid.jpg}}}$ \and 
M. Gillon\inst{31} $^{\href{https://orcid.org/0000-0003-1462-7739}{\includegraphics[scale=0.5]{figures/orcid.jpg}}}$ \and 
M. Güdel\inst{37} $^{\href{https://orcid.org/0000-0001-9818-0588}{\includegraphics[scale=0.5]{figures/orcid.jpg}}}$ \and 
S. Hoyer\inst{4} $^{\href{https://orcid.org/0000-0003-3477-2466}{\includegraphics[scale=0.5]{figures/orcid.jpg}}}$ \and 
K. G. Isaak\inst{38} $^{\href{https://orcid.org/0000-0001-8585-1717}{\includegraphics[scale=0.5]{figures/orcid.jpg}}}$ \and 
L. L. Kiss\inst{39,40} $^{\href{https://orcid.org/0000-0002-3234-1374}{\includegraphics[scale=0.5]{figures/orcid.jpg}}}$ \and 
E. Kopp\inst{41,42,43} \and
J. Laskar\inst{44} $^{\href{https://orcid.org/0000-0003-2634-789X}{\includegraphics[scale=0.5]{figures/orcid.jpg}}}$ \and 
A. Lecavelier des Etangs\inst{45} $^{\href{https://orcid.org/0000-0002-5637-5253}{\includegraphics[scale=0.5]{figures/orcid.jpg}}}$ \and 
C. Lovis\inst{2} $^{\href{https://orcid.org/0000-0001-7120-5837}{\includegraphics[scale=0.5]{figures/orcid.jpg}}}$ \and 
D. Magrin\inst{24} $^{\href{https://orcid.org/0000-0003-0312-313X}{\includegraphics[scale=0.5]{figures/orcid.jpg}}}$ \and 
P. F. L. Maxted\inst{46} $^{\href{https://orcid.org/0000-0003-3794-1317}{\includegraphics[scale=0.5]{figures/orcid.jpg}}}$ \and 
C. Mordasini\inst{14,47} $^{\href{https://orcid.org/0000-0002-1013-2811}{\includegraphics[scale=0.5]{figures/orcid.jpg}}}$ \and 
V. Nascimbeni\inst{24} $^{\href{https://orcid.org/0000-0001-9770-1214}{\includegraphics[scale=0.5]{figures/orcid.jpg}}}$ \and 
G. Olofsson\inst{3} $^{\href{https://orcid.org/0000-0003-3747-7120}{\includegraphics[scale=0.5]{figures/orcid.jpg}}}$ \and 
R. Ottensamer\inst{48} $^{\href{https://orcid.org/0000-0001-5684-5836}{\includegraphics[scale=0.5]{figures/orcid.jpg}}}$ \and 
I. Pagano\inst{49} $^{\href{https://orcid.org/0000-0001-9573-4928}{\includegraphics[scale=0.5]{figures/orcid.jpg}}}$ \and 
E. Pallé\inst{16} $^{\href{https://orcid.org/0000-0003-0987-1593}{\includegraphics[scale=0.5]{figures/orcid.jpg}}}$ \and 
G. Peter\inst{41} $^{\href{https://orcid.org/0000-0001-6101-2513}{\includegraphics[scale=0.5]{figures/orcid.jpg}}}$ \and 
G. Piotto\inst{24,50} $^{\href{https://orcid.org/0000-0002-9937-6387}{\includegraphics[scale=0.5]{figures/orcid.jpg}}}$ \and 
D. Pollacco\inst{10} \and
R. Ragazzoni\inst{24,50} $^{\href{https://orcid.org/0000-0002-7697-5555}{\includegraphics[scale=0.5]{figures/orcid.jpg}}}$ \and 
N. Rando\inst{28} \and
H. Rauer\inst{26,42,43} $^{\href{https://orcid.org/0000-0002-6510-1828}{\includegraphics[scale=0.5]{figures/orcid.jpg}}}$ \and 
I. Ribas\inst{18,19} $^{\href{https://orcid.org/0000-0002-6689-0312}{\includegraphics[scale=0.5]{figures/orcid.jpg}}}$ \and 
N. C. Santos\inst{12,13} $^{\href{https://orcid.org/0000-0003-4422-2919}{\includegraphics[scale=0.5]{figures/orcid.jpg}}}$ \and 
G. Scandariato\inst{49} $^{\href{https://orcid.org/0000-0003-2029-0626}{\includegraphics[scale=0.5]{figures/orcid.jpg}}}$ \and 
D. Ségransan\inst{2} $^{\href{https://orcid.org/0000-0003-2355-8034}{\includegraphics[scale=0.5]{figures/orcid.jpg}}}$ \and 
A. E. Simon\inst{15} $^{\href{https://orcid.org/0000-0001-9773-2600}{\includegraphics[scale=0.5]{figures/orcid.jpg}}}$ \and 
A. M. S. Smith\inst{26} $^{\href{https://orcid.org/0000-0002-2386-4341}{\includegraphics[scale=0.5]{figures/orcid.jpg}}}$ \and 
M. Steller\inst{1} $^{\href{https://orcid.org/0000-0003-2459-6155}{\includegraphics[scale=0.5]{figures/orcid.jpg}}}$ \and 
Gy. M. Szabó\inst{51,52} \and
N. Thomas\inst{15} $^{\href{https://orcid.org/0000-0002-0146-0071}{\includegraphics[scale=0.5]{figures/orcid.jpg}}}$ \and 
S. Udry\inst{2} $^{\href{https://orcid.org/0000-0001-7576-6236}{\includegraphics[scale=0.5]{figures/orcid.jpg}}}$ \and 
B. Ulmer\inst{41,42,43} \and
V. Van Grootel\inst{32} $^{\href{https://orcid.org/0000-0003-2144-4316}{\includegraphics[scale=0.5]{figures/orcid.jpg}}}$ \and 
J. Venturini\inst{2} $^{\href{https://orcid.org/0000-0001-9527-2903}{\includegraphics[scale=0.5]{figures/orcid.jpg}}}$ \and 
N. A. Walton\inst{53} $^{\href{https://orcid.org/0000-0003-3983-8778}{\includegraphics[scale=0.5]{figures/orcid.jpg}}}$
}

\authorrunning{A. F. Krenn et al.}

\institute{Space Research Institute, Austrian Academy of Sciences, Schmiedlstrasse 6, A-8042 Graz, Austria\\ 
\email{andreas.krenn@oeaw.ac.at} \\
\and
Observatoire Astronomique de l'Université de Genève, Chemin Pegasi 51, CH-1290 Versoix, Switzerland \\
\and
Department of Astronomy, Stockholm University, AlbaNova University Center, 10691 Stockholm, Sweden \\
\and
Aix Marseille Univ, CNRS, CNES, LAM, 38 rue Frédéric Joliot-Curie, 13388 Marseille, France \\
\and
Centre for Exoplanet Science, SUPA School of Physics and Astronomy, University of St Andrews, North Haugh, St Andrews KY16 9SS, UK \\
\and
Division Technique INSU, CS20330, 83507 La Seyne sur Mer cedex, France \\
\and
ETH Zurich, Department of Physics, Wolfgang-Pauli-Strasse 2, CH-8093 Zurich, Switzerland \\
\and
Cavendish Laboratory, JJ Thomson Avenue, Cambridge CB3 0HE, UK \\
\and
Ludwig Maximilian University, University Observatory Munich, Scheinerstrasse 1, Munich D-81679, Germany \\
\and
Department of Physics, University of Warwick, Gibbet Hill Road, Coventry CV4 7AL, United Kingdom \\
\and
University of Bern, ARTORG Center for Biomedical Engineering Research, Murtenstrasse 50, CH-3008, Bern, Switzerland \\
\and
Instituto de Astrofisica e Ciencias do Espaco, Universidade do Porto, CAUP, Rua das Estrelas, 4150-762 Porto, Portugal \\
\and
Departamento de Fisica e Astronomia, Faculdade de Ciencias, Universidade do Porto, Rua do Campo Alegre, 4169-007 Porto, Portugal \\ 
\and
Center for Space and Habitability, University of Bern, Gesellschaftsstrasse 6, 3012 Bern, Switzerland \\
\and
Physikalisches Institut, University of Bern, Sidlerstrasse 5, 3012 Bern, Switzerland \\
\and
Instituto de Astrofisica de Canarias, 38200 La Laguna, Tenerife, Spain \\
\and
Departamento de Astrofisica, Universidad de La Laguna, 38206 La Laguna, Tenerife, Spain \\
\and
Institut de Ciencies de l'Espai (ICE, CSIC), Campus UAB, Can Magrans s/n, 08193 Bellaterra, Spain \\
\and
Institut d'Estudis Espacials de Catalunya (IEEC), 08034 Barcelona, Spain \\
\and
Admatis, 5. Kandó Kálmán Street, 3534 Miskolc, Hungary \\
\and
Depto. de Astrofisica, Centro de Astrobiologia (CSIC-INTA), ESAC campus, 28692 Villanueva de la Cañada (Madrid), Spain \\
\and
Almatech SA, EPFL Innovation Park, Bâtiment D, 1015 Lausanne, Switzerland \\
\and
Université Grenoble Alpes, CNRS, IPAG, 38000 Grenoble, France \\
\and
INAF, Osservatorio Astronomico di Padova, Vicolo dell'Osservatorio 5, 35122 Padova, Italy \\
\and
Center for Space and Habitability, Gesellsschaftstrasse 6, 3012 Bern, Switzerland \\
\and
Institute of Planetary Research, German Aerospace Center (DLR), Rutherfordstrasse 2, 12489 Berlin, Germany \\ 
\and
Université de Paris, Institut de physique du globe de Paris, CNRS, F-75005 Paris, France \\
\and
ESTEC, European Space Agency, 2201AZ, Noordwijk, NL \\
\and
INAF, Osservatorio Astrofisico di Torino, Via Osservatorio, 20, I-10025 Pino Torinese To, Italy \\
\and
Centre for Mathematical Sciences, Lund University, Box 118, 221 00 Lund, Sweden \\
\and
Astrobiology Research Unit, Université de Liège, Allée du 6 Août 19C, B-4000 Liège, Belgium \\
\and
Space sciences, Technologies and Astrophysics Research (STAR) Institute, Université de Liège, Allée du 6 Août 19C, 4000 Liège, Belgium \\
\and
Centre Vie dans l’Univers, Faculté des sciences, Université de Gen\`eve, Quai Ernest-Ansermet 30, CH-1211 Gen\`eve 4, Switzerland \\
\and
Leiden Observatory, University of Leiden, PO Box 9513, 2300 RA Leiden, The Netherlands \\
\and
Department of Space, Earth and Environment, Chalmers University of Technology, Onsala Space Observatory, 439 92 Onsala, Sweden \\ 
\and
Dipartimento di Fisica, Universita degli Studi di Torino, via Pietro Giuria 1, I-10125, Torino, Italy \\
\and
University of Vienna, Department of Astrophysics, Türkenschanzstrasse 17, 1180 Vienna, Austria \\
\and
Science and Operations Department - Science Division (SCI-SC), Directorate of Science, European Space Agency (ESA), European Space Research and Technology Centre (ESTEC), Keplerlaan 1, 2201-AZ Noordwijk, The Netherlands \\
\and
Konkoly Observatory, Research Centre for Astronomy and Earth Sciences, 1121 Budapest, Konkoly Thege Miklós út 15-17, Hungary \\
\and
ELTE E\"otv\"os Lor\'and University, Institute of Physics, P\'azm\'any P\'eter s\'et\'any 1/A, 1117 Budapest, Hungary \\
\and
Institute of Optical Sensor Systems, German Aerospace Center (DLR), Rutherfordstrasse 2, 12489 Berlin, Germany \\
\and
Zentrum für Astronomie und Astrophysik, Technische Universität Berlin, Hardenbergstr. 36, D-10623 Berlin, Germany \\
\and
Institut für Geologische Wissenschaften, Freie Universität Berlin, 12249 Berlin, Germany \\
\and
IMCCE, UMR8028 CNRS, Observatoire de Paris, PSL Univ., Sorbonne Univ., 77 av. Denfert-Rochereau, 75014 Paris, France \\
\and
Institut d'astrophysique de Paris, UMR7095 CNRS, Université Pierre \& Marie Curie, 98bis blvd. Arago, 75014 Paris, France \\
\and
Astrophysics Group, Keele University, Staffordshire, ST5 5BG, United Kingdom \\
\and
Physikalisches Institut, University of Bern, Gesellschaftsstrasse 6, 3012 Bern, Switzerland \\
\and
Department of Astrophysics, University of Vienna, Tuerkenschanzstrasse 17, 1180 Vienna, Austria \\ 
\and
INAF, Osservatorio Astrofisico di Catania, Via S. Sofia 78, 95123 Catania, Italy \\
\and
Dipartimento di Fisica e Astronomia "Galileo Galilei", Universita degli Studi di Padova, Vicolo dell'Osservatorio 3, 35122 Padova, Italy \\
\and
ELTE E\"otv\"os Lor\'and University, Gothard Astrophysical Observatory, 9700 Szombathely, Szent Imre h. u. 112, Hungary \\
\and
MTA-ELTE Exoplanet Research Group, 9700 Szombathely, Szent Imre h. u. 112, Hungary \\
\and
Institute of Astronomy, University of Cambridge, Madingley Road, Cambridge, CB3 0HA, United Kingdom 
}

\date{Received 20.09.2022; accepted 15.12.2022}

 
  \abstract
   {Measurements of the occultation of an exoplanet at visible wavelengths allow us to determine the reflective properties of a planetary atmosphere. The observed occultation depth can be translated into a geometric albedo. This in turn aids in characterising the structure and composition of an atmosphere by providing additional information on the wavelength-dependent reflective qualities of the aerosols in the atmosphere.}
   {Our aim is to provide a precise measurement of the geometric albedo of the gas giant HD\,189733b by measuring the occultation depth in the broad optical bandpass of CHEOPS (350 -- 1100 nm).}
   {We analysed 13 observations of the occultation of HD\,189733b performed by CHEOPS utilising the \texttt{Python} package \texttt{PyCHEOPS}. The resulting occultation depth is then used to infer the geometric albedo accounting for the contribution of thermal emission from the planet. We also aid the analysis by refining the transit parameters combining observations made by the TESS and CHEOPS space telescopes.}
   {We report the detection of an $24.7 \pm 4.5$ ppm occultation in the CHEOPS observations. This occultation depth corresponds to a geometric albedo of $0.076 \pm 0.016$. Our measurement is consistent with models assuming the atmosphere of the planet to be cloud-free at the scattering level and absorption in the CHEOPS band to be dominated by the resonant Na doublet. Taking into account previous optical-light occultation observations obtained with the Hubble Space Telescope, both measurements combined are consistent with a super-stellar Na elemental abundance in the dayside atmosphere of HD\,189733b. We further constrain the planetary Bond albedo to between 0.013 and 0.42 at 3$\sigma$ confidence.}
   {We find that the reflective properties of the  HD\,189733b dayside atmosphere are consistent with a cloud-free atmosphere having a super-stellar metal content. When compared to an analogous CHEOPS measurement for HD\,209458b, our data hint at a slightly lower geometric albedo for HD\,189733b ($0.076 \pm 0.016$) than for HD\,209458b ($0.096 \pm 0.016$), or a higher atmospheric Na content in the same modelling framework. While our constraint on the Bond albedo is consistent with previously published values, we note that the higher-end values of $\sim$0.4, as derived previously from infrared phase curves, would also require peculiarly high reflectance in the infrared, which again would make it more difficult to  disentangle reflected and emitted light in the total observed flux, and therefore to correctly account for reflected light in the interpretation of those phase curves. Lower reported values for the Bond albedos are less affected by this ambiguity.}


\maketitle
%

\begin{table*}
    \caption{Overview of the log details of CHEOPS observations.}
    \centering
    \begin{tabular}{cccc}
    \toprule
    \toprule
    Visit\# & Start Date & End Date & File Key \\
    \midrule
    &&Transit observations &\\
    1 & 2021-08-10 23:46 & 2021-08-11 08:39 & CH\textunderscore PR100019\textunderscore TG000201\textunderscore V0200 \\
    2 & 2021-08-30 20:00 & 2021-08-31 04:53 & CH\textunderscore PR100019\textunderscore TG000301\textunderscore V0200 \\
    \midrule
    &&Occultation observations & \\
    1 & 2021-06-30 21:30 & 2021-07-01 07:55 & CH\textunderscore PR100016\textunderscore TG013601\textunderscore V0200 \\
    2 & 2021-07-07 11:47 & 2021-07-07 21:39 & CH\textunderscore PR100016\textunderscore TG013602\textunderscore V0200 \\
    3 & 2021-07-09 16:56 & 2021-07-10 02:52 & CH\textunderscore PR100016\textunderscore TG013603\textunderscore V0200 \\
    4 & 2021-07-11 22:39 & 2021-07-12 08:07 & CH\textunderscore PR100016\textunderscore TG013604\textunderscore V0200 \\
    5 & 2021-07-16 09:11 & 2021-07-16 19:08 & CH\textunderscore PR100016\textunderscore TG013605\textunderscore V0200 \\
    6 & 2021-07-20 19:43 & 2021-07-21 05:39 & CH\textunderscore PR100016\textunderscore TG013606\textunderscore V0200 \\
    7 & 2021-07-23 01:53 & 2021-07-23 11:49 & CH\textunderscore PR100016\textunderscore TG013607\textunderscore V0200 \\
    8 & 2021-08-05 08:08 & 2021-08-05 17:28 & CH\textunderscore PR100016\textunderscore TG013608\textunderscore V0200 \\
    9 & 2021-08-14 04:52 & 2021-08-14 16:17 & CH\textunderscore PR100016\textunderscore TG013609\textunderscore V0200 \\
    10 & 2021-08-16 11:09 & 2021-08-16 20:32 & CH\textunderscore PR100016\textunderscore TG013610\textunderscore V0200 \\
    11 & 2022-07-10 19:35 & 2022-07-11 05:32 & CH\textunderscore PR100016\textunderscore TG016601\textunderscore V0200 \\
    12 & 2022-07-17 11:13 & 2022-07-17 20:31 & CH\textunderscore PR100016\textunderscore TG016602\textunderscore V0200 \\
    13 & 2022-08-02 00:18 & 2022-08-02 10:58 & CH\textunderscore PR100016\textunderscore TG016603\textunderscore V0200 \\
    \bottomrule
    \end{tabular}
    \tablefoot{Time notation follows the ISO-8601 convention. The file keys can be used to retrieve data from the CHEOPS archive.}
    \label{tab:file_keys}
\end{table*}

\section{Introduction}
\label{sec_introduction}
The reflective properties of an exoplanetary atmosphere are quantified by the geometric albedo \citep{Russel1916}, which is defined as the albedo of the planet at full phase. The global energy budget of a planet is dependent on how much light from the host star is able to enter the atmosphere without being reflected at its top. The reflectivity of the atmosphere can also be wavelength-dependent as different atmospheric constituents absorb and reflect light at different wavelengths differently. This results in the geometric albedo being dependent on the spectral range of the observed light \citep{Sudarsky2000,Heng2013,Parmentier2016}. The measurement of the albedo therefore provides an additional observational constraint when modelling the atmospheric structure and composition of an exoplanet.

In practice, the geometric albedo can be determined by observing the secondary eclipse (occultation) of the planet at optical wavelengths. While observations of the occultation at infrared wavelengths are limited to the thermal contribution of the planetary emission, measurements at optical wavelengths also allow for the characterisation of the reflective contribution. The total occulted flux is a combination of thermal emission from the planet and reflected stellar light. By accounting for the thermal contribution to the total occulted flux \citep[e.g.][]{Cowan2011,Heng2013,Wong2020,Wong2021}, the geometric albedo can then be inferred from the  remaining reflective contribution.

Such measurements have   been done with a variety of   space telescopes. Both \citet{Angerhausen2015} and \citet{Esteves2015} report geometric albedos for 20 and 14 planets, respectively. They determined the albedos with phase curve observations performed by the Kepler Space telescope. Both studies find that geometric albedos of gas giants in the Kepler bandpass \citep[$420$ - $910$ nm;][]{Koch2010} are usually   $<0.1$.

These findings are also supported by \citet{Heng2013}, who obtained $A_g < 0.15$ for 9 out of 11 planets looked at in their study. The two exceptions are HAT-P-7b and Kepler-7b, for which they report $A_g = 0.225 \pm 0.004$ and $A_g = 0.352 \pm 0.023$, respectively. The case of the high albedo of Kepler-7b, however,  has been widely discussed, and \citet{Heng2021} revised the value to $A_g = 0.25^{+0.01}_{-0.02}$. The high albedos for these planets are attributed to the presence of clouds or condensates in the atmosphere. All of these observations also prove the capability of space-grade photometry to detect low amplitude occultation signals with, for example, a measured occultation depth of $10.9 \pm 2.2$ ppm for TrES-2b and $16.5 \pm 4.5$ ppm for Kepler-8b \citep{Kipping2011,Angerhausen2015}.

\citet{Wong2020} and \citet{Wong2021} report detections of secondary eclipses at optical wavelengths for 15 planets from data acquired with the Transiting Exoplanet Survey Satellite (TESS) \citep{Ricker2014}. Due to design and purpose, the photometric precision of TESS is smaller than that of Kepler. The smallest occultation depth detected with at least 3$\sigma$ confidence in these studies is that of WASP-100b at $94\pm17$ ppm \citep{Wong2021}. However, the reported uncertainties are often an order of magnitude higher for other systems. They also infer geometric albedos for the planets with a detected occultation, while being careful to account for the thermal contribution to the total occulted flux. For planets with dayside temperatures below $1500$ K, they find that with increasing temperature the geometric albedo decreases because fewer condensates are created in the atmosphere. However, for planets with dayside temperatures between $1500$ and $3000$ K, they find a weak positive correlation of the dayside brightness temperature and the geometric albedo. This means that the reflective qualities of planets with very high temperatures improve again despite the temperature increase. They attribute this to high-temperature condensates, which cause higher atmospheric reflectivity. Alternatively they also suggest that opacity sources may contribute additional unaccounted emission at visible wavelengths \citep{Cowan2011}, which lead to biased high albedos. 

\begin{figure}
    \centering
    \includegraphics[width=\hsize]{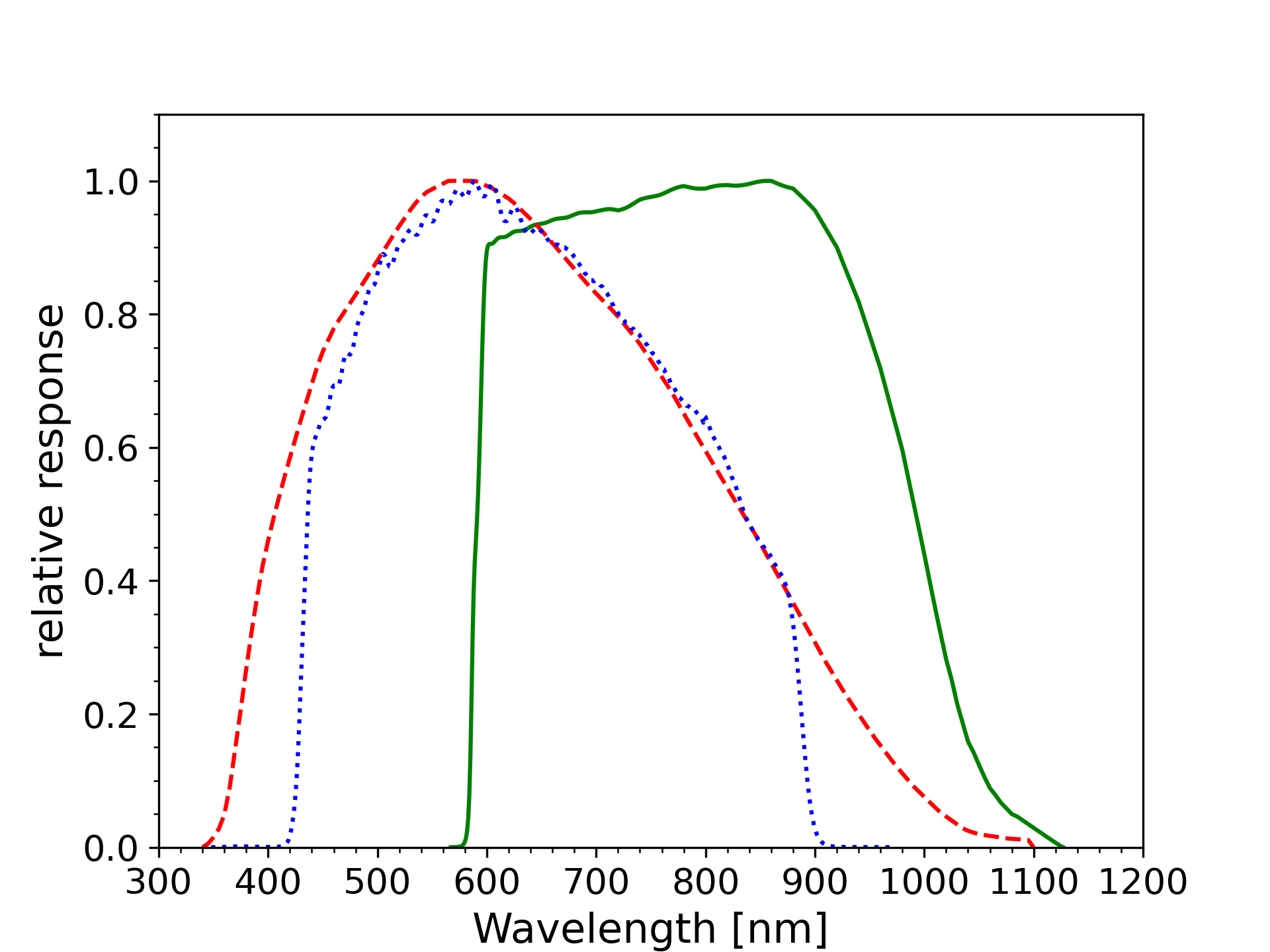}
    \caption{Wavelength-dependent response functions of CHEOPS (dashed red), Kepler (dotted blue), and TESS (solid green). The response functions are from the Spanish Virtual Observatory Filter Profile Service \citep{svo1,svo2}.}
    \label{fig:bandpass}
\end{figure}

The Characterising Exoplanet Satellite \citep[CHEOPS;][]{Benz2021} has been used to detect occultations of several planets. \citet{Lendl2020} reported the first detection of an occultation (WASP-189b) with CHEOPS. \citet{Brandeker2022} used CHEOPS to observe occultations of the hot Jupiter HD\,209458b at optical wavelengths. They prove the capability of this space telescope to also detect shallow secondary eclipses by reporting a detection of an occultation in the CHEOPS bandpass with a depth of $20.4^{+3.2}_{-3.3}$ ppm. Apart from the higher photometric precision, CHEOPS also provides a wider bandpass ($350$ - $1100$ nm) than TESS ($580$ - $1120$ nm). The response functions for CHEOPS, TESS, and Kepler are shown in Figure \ref{fig:bandpass}. The CHEOPS and Kepler bandpasses are very similar. The TESS bandpass does not cover the bluer parts of the visible spectrum, but is much more sensitive at infrared wavelengths and therefore to thermal emission from planets, while Kepler and CHEOPS are more sensitive to reflection.

\citet{Bouchy2005} reported the discovery of the hot Jupiter HD\,189733b, which was among the first exoplanets for which both radial velocity and photometric transit measurements became available. The combination of these two complementary methods allowed for a first characterisation of the planet using its density. It has since become a benchmark for studying hot Jupiters and their atmospheres. \citet{Deming2006} were the first to report the detection of an occultation of HD\,189733b in data obtained with the Spitzer Space Telescope. \citet{Knutson2007} and \citet{Knutson2009} used phase curves observed by Spitzer in the 8 and 24 $\mu$m channels to study the day--night contrast of the planet. They reported a maximum brightness temperature of 1212 $\pm$ 11 $K$ at a wavelength of 8 $\mu$m. They also find a small day--night contrast and a bright-spot offset in both wavelength channels, indicating that the energy absorbed at the day-side of the planet is efficiently redistributed across the atmosphere. \citet{Agol2010} used a total of seven occultation observations by Spitzer to study phase variations of the infrared flux emitted by the planet. Most notably they reported no significant eclipse depth variations over the two-year baseline of their observations. The set of Spitzer phase curve and occultation observations was complemented by additional observations in the 3.6 and 4.5 $\mu$m channels \citep{Knutson2012}. They also reported no significant eclipse depth variations and present enhanced absorption in the 4.5 $\mu$m channel as evidence for the existence of vertical mixing resulting in excess of CO.

Thermal phase curves can be used to derive estimates of the Bond albedo of a planet by measuring day- and night-side temperatures and computing the equilibrium temperature of the planet \citep{Cowan2011,Schwartz2015}. The Bond albedo is a measure of the total reflectance of a planet across all wavelengths and in all directions. It is specifically important when determining the energy budget of a planet. This has been done several times for the case of HD\,189733b using the Spitzer thermal phase curve observations. \citet{Schwartz2015} and \citet{Schwartz2017} report consistent measurements of $A_B = 0.37_{-0.05}^{+0.04}$ and $A_B = 0.41 \pm 0.07$ respectively. Notably, \citet{Zhang2018} show that applying the same approach used in \citet{Schwartz2017} to only Spitzer 3.6 and 4.5 $\mu$m phase curves (discarding the 8 and 24 $\mu$m phase curves), yields a lower value (but still consistent within $1\sigma$) of $A_B \approx 0.3 \pm 0.1$. All of these works use similar approaches (error weighted means) to convert  broad-band brightness temperatures to day- and nightside effective temperatures. \citet{Keating2019}, on the other hand, retrieve a significantly lower estimate of $A_B = 0.16_{-0.10}^{+0.11}$ from the same  datasets using a Gaussian process regression developed by \citet{Pass2019} to create a wavelength-dependent brightness map, which in turn is used to estimate effective temperatures.

\citet{Evans2013} studied the occultation of HD\,189733b at optical wavelengths using observations by the Hubble Space Telescope (HST). They were able to detect the occultation signal in the 290 -- 450 nm wavelength channel at $126^{+37}_{-36}$ ppm, but were unable to detect it in the redder 450 -- 570 nm wavelength channel where the signal is expected to be significantly smaller. They provide an upper limit of the geometric albedo of $A_g < 0.12$ across 450 -- 570 nm at 1$\sigma$ confidence, while determining a geometric albedo across 290 -- 450 nm of $A_g = 0.40 \pm 0.12$. They interpreted the decrease in  the albedo towards longer wavelengths as an indication of the presence of optically thick reflective clouds on the dayside. In particular, they suspect that  sodium absorption suppresses the scattering of light from the red part of the visible spectrum.

Using the 390 -- 435 nm and 435--480 nm channels of the HST data published in \citet{Evans2013}, \citet{Wiktorowicz2015} computed a geometric albedo of HD\,189733b in the B-band (390 -- 480 nm) of $A_g = 0.226 \pm 0.091$. They supported their computation with polarimetric observations, which allowed for the determination of a 3$\sigma$ confidence upper limit of $A_g < 0.37$ in the B-band (in this case 391--482 nm). Combining these photometric and polarimetric observations they reject a B-band geometric albedo of $A_g = 0.61 \pm 0.12$, which was previously reported by \citet{Berdyugina2011} using different polarimetric observations. Their result is consistent with models including the presence of Rayleigh scattering in the atmosphere.  

In this work we present observations of the occultation of HD\,189733b at optical wavelengths performed by CHEOPS with the aim of utilising the exquisite precision of the space telescope for bright stars to determine the geometric albedo of the planet. Additionally, we also present transit observations of the target performed by TESS and CHEOPS that aid in   constraining the planetary parameters in the course of the analysis of the occultation observations. Section \ref{sec_data} contains a description of the acquired data, and in Section \ref{sec_analysis} the analysis of the data is described in detail. In Section \ref{sec_atmos}, using the retrieved occultation depth, we provide an estimate of the geometric albedo of the planet and discuss it in the context of other geometric albedo measurements of similar planets. Finally, we also derive lower and upper boundaries for the Bond albedo of HD\,189733b and compare them with previous measurements of the Bond albedo from thermal phase curves.


\section{Description of acquired data}
\label{sec_data}

\subsection{Transit data}
\label{subsec_transit_data}

HD\,189733 was observed by TESS in Sector 41 of cycle 4 of the extended mission at 2 min cadence. In our analysis we used the \texttt{Pre-search Data Conditioning Simple Aperture Photometry} (PDC-SAP) provided by the Science Processing Operations Center (SPOC) pipeline \citep{Smith2012,Stumpe2014,2016SPIE.9913E..3EJ}. CHEOPS also performed two transit observations of the target in August 2021 (see Table \ref{tab:file_keys}). The CHEOPS observations are available as two different data products \citep{Benz2021}: sub-arrays, which contain a circular region around the target with a radius of 100 pixels and are a product of combining three individual 10.45 s exposures resulting in an effective cadence of 31.35 s, and imagettes, which contain circular subsections of a 25-pixel radius around the target and are available at a cadence equal to the exposure time of 10.45 s. Aperture photometry is available only for the sub-arrays via the official CHEOPS Data Reduction Pipeline \citep[DRP;][]{Hoyer2020}. However, PSF photometry can be performed on the imagettes using PIPE\footnote{\url{http://github.com/alphapsa/PIPE}}\citep[see also][]{2021A&A...651L..12M, 2021A&A...654A.159S, Brandeker2022}. 

We processed all of the CHEOPS transit observations with the DRP using an aperture radius of 30 pixels (\emph{RSUP} aperture radius option). This option was chosen as it   ensures that the whole PSF is   included in the aperture radius, and minimises the number of nearby contaminating stars. Notably, we   also performed both the transit and occultation analysis using data processed with PIPE, and achieved consistent results with both reduction alternatives.

In general, CHEOPS observations are affected by instrumental noise such as stray light from the Earth and the Moon (Moon glint), smearing effects, or spacecraft jitter. The flux measurements show a particularly strong correlation with the spacecraft roll angle \citep[see also][]{Lendl2020,Bonfanti2021}. The spacecraft is designed to rotate around itself exactly once every orbit. Therefore, the roll angle parameter is directly linked to the orbital position of the spacecraft. Instrumental noise must be accounted for during the data analysis in order to identify the transit and occultation signals of the planet (see Section \ref{sec_analysis}). Prior to performing the transit analysis, we removed all of those points that were flagged by the DRP; this includes those points that are contaminated, for example by  cosmic rays. Additionally, we performed a sigma clipping and removed all points with the median absolute deviation (MAD) higher than 5 to discard outliers.

\subsection{Occultation data}
\label{subsec_occultation_data}

CHEOPS observed HD\,189733 ten times in a time period from   30 June 2021 until  16   August 2021, and three times from   10   July 2022 until   2   August 2022 (see  Table \ref{tab:file_keys}). Each visit contains a single occultation event at the middle of the observation with out-of-occultation data being acquired both before and after the occultation. Each individual visit   comprises either six or seven CHEOPS orbits. A single CHEOPS orbit covers roughly 100 minutes. However, the target cannot be observed throughout the entire orbit, due to Earth occultations and South Atlantic Anomaly (SAA) crossings. This leads to gaps in the observations, with a width that varies from visit to visit, spanning values  between 20 and 40 minutes, depending on target coordinates and observation time.

For the occultation observations, the dataset of each individual visit was also processed with the standard CHEOPS DRP version 13.1.0 \citep{Hoyer2020}. Again an aperture radius of 30 pixels (\emph{RSUP} aperture radius option) was used during processing. To remove outliers, we applied a MAD clipping with a clipping factor of $\sigma = 2.5$ on all datasets. Since the observed roll angle trends in the data are partially caused by changes in the amount of background light, we opted for a clipping of data points with high background values. All points for which the background estimate $> 700$ ADU were removed. This procedure  especially removes data points shortly before and after an Earth occultation, when stray light from Earth's atmosphere hits the telescope, as well as data points with increased flux levels due to a crossing of the SAA. Finally, we also removed data points with a smear estimate larger than $9 \times 10^{-5}$ times the median flux value of the whole dataset as well as all data points with at least one centroid coordinate being shifted more than $1$ pixel from the median centroid position. Both the threshold for background clipping and the threshold for smear estimate clipping were determined by inspecting plots showing the correlation of the corresponding parameters with the observed fluxes. The thresholds were chosen in such a way that they represent a trade-off between ensuring that data points  affected very strongly by instrumental systematics are   clipped and minimising the number of clipped points.


\section{Data analysis}
\label{sec_analysis}

\begin{table}
\caption{Adopted stellar parameters of HD\,189733}             
\centering                          
\begin{tabular}{cccc}        
\hline\hline                 
Parameter & Value & Unit & Source \\    
\hline                   
Vmag & $7.648$ & - & SWEET-Cat \\
Spectral Type & K2 V & - & SWEET-Cat \\
$T_{\mathrm{eff} eff}$ & $4969\pm43$ & K & SWEET-Cat \\
 $\log g$  & $4.51\pm0.03$ & - & GAIA eDR3 \\
Radius & $0.784\pm0.007$ & $R_{\odot}$ & This work \\
Mass & $0.783\pm0.041$ & $M_{\odot}$ & This work \\
Density &$1.6246\pm 0.0936$ &  $\rho_\odot$ & This work \\
Age &  $8.3_{-3.1}^{+3.6}$ & Gyr & This work \\
$\left[\mathrm{Fe/H}\right]$ & $-0.07\pm0.02$ & - & SWEET-Cat \\
$\left[\mathrm{C/H}\right]$ & $-0.09\pm0.11$ & - & This work \\
$\left[\mathrm{N/H}\right]$ & $-0.05\pm0.16$ & - & This work \\
$\left[\mathrm{O/H}\right]$ & $-0.00\pm0.11$ & - & This work \\
$\left[\mathrm{Na/H}\right]$ & $-0.05\pm0.02$ & - & This work \\
\hline                               
\end{tabular}
\label{table_stellar}      
\tablefoot{The Gaia ID of HD\,189733b is 1827242816201846144}
\end{table}

\subsection{Stellar properties}
\label{subsec_stellar}
To aid  our transit and   occultation analyses, we provided a prior on the stellar density of the host star HD\,189733. The stellar density was calculated from stellar radius and mass estimates computed specifically for this work. To derive the radius we used a modified Markov chain Monte Carlo (MCMC) infrared flux method (IRFM; \citealt{Blackwell1977,Schanche2020}) with spectral priors taken from values available in the SWEET-Cat catalogue (\citealt{Sousa2021}; for more details see \citealt{Wilson2022}). To retrieve a mass estimate we assumed the stellar $T_{\mathrm{eff}}$, [Fe/H], and radius as the basic set of input parameters to derive the isochronal mass and age values from two different stellar evolutionary models. To retrieve the first pair of mass and age values we used the interpolation capability of the isochrone placement algorithm \citep{bonfanti15,bonfanti16} to fit the input parameters within pre-computed grids of PARSEC\footnote{ \textsl{PA}dova and T\textsl{R}ieste \textsl{S}tellar \textsl{E}volutionary \textsl{C}ode: \url{http://stev.oapd.inaf.it/cgi-bin/cmd}} v1.2S \citep{marigo17} isochrones and tracks. In particular, we added $v\sin{i}=3.5\pm1.0$ km/s \citep{Bouchy2005} to the basic set of input parameters because the synergy between isochrones and gyrochronology improves the routine convergence, as detailed in \citet{bonfanti16}. A second pair of mass and age values was computed by the Code Liègeois d'Évolution Stellaire  \citep[CLES;][]{scuflaire08}, which generates the best-fit stellar track according to the basic set of input parameters following the Levenberg-Marquadt minimisation scheme \citep{salmon21}.
As discussed in \citet{Bonfanti2021}, we finally checked the mutual consistency of the two respective pairs of outcomes through a $\chi^2$-based criterion, and we merged the mass and age distributions to obtain the results used in this study. All adopted stellar parameters including the stellar density are presented in Table~\ref{table_stellar}.

To aid our interpretation of the measured geometric albedo we also determined the stellar sodium abundance from the  publicly available optical spectrum obtained with the High Accuracy Radial velocity Planet Searcher (HARPS) spectrograph using the derived and adopted stellar parameters. We adopted the classical curve-of-growth analysis method assuming local thermodynamic equilibrium. We used the ARES v2 code\footnote{The last version of ARES code (ARES v2) can be downloaded at http://www.astro.up.pt/$\sim$sousasag/ares} \citep{Sousa-15} to measure the equivalent widths of the spectral lines. Then we used a grid of Kurucz model atmospheres \citep{Kurucz-93} and the radiative transfer code MOOG \citep{Sneden-73} to convert the EWs into abundances closely following the methods described in  \citet{Adibekyan-12} and \citet{Adibekyan-15}, among others.

Given the low temperature of the star, it was not possible to determine the abundances of C, N, and O directly from the HARPS spectrum. The abundances of these elements have been estimated using the available abundance datasets of solar neighbourhood stars: \citet{DelgadoMena-21} for C, \citet{SuarezAndres-16} for N, and \citet{BertranDeLis-15} for O. We determined the mean value and standard deviation of the abundances of these elements for stars with metallicities similar  to that of HD\,189733 ([Fe/H] = -0.07$\pm$0.02 dex). Expanding the metallicity range by  factors of 5 ([Fe/H] = -0.07$\pm$0.10 dex) and 10 ([Fe/H] = -0.07$\pm$0.20 dex) has a minor impact on the mean abundance and its standard deviation. Finally, to transform the relative atmospheric abundances (see Table \ref{table_stellar}) into absolute abundances we adopted the solar reference values from \cite{Asplund21}. This yielded the following elemental abundances of carbon, oxygen, nitrogen, and sodium: C/H = ($2.4\pm0.7) \times 10^{-4}$, (O/H = $6.6\pm2.7) \times 10^{-5}$, C/H = ($5.1\pm1.3) \times 10^{-4}$, Na/H = ($1.6\pm0.2) \times 10^{-6}$. These values are   used as input to our computation of the planetary geometric albedo (see Sect. \ref{sec_albedo}).

\subsection{Transit analysis}
\label{subsec_transit_analysis}

\begin{table*}
    \caption{Retrieved planetary and stellar parameters}
    \centering
    \begin{tabular}{lcccr}
    \toprule
    \toprule
    Parameters & Symbols & Priors & Values &  Units \\
    \midrule
    Planetary parameters&&&& \\
    \quad Orbital period & $P$ &$\mathcal{N}(2.2185752,7.7 \times 10^{-8})$ & {\small $2.2185751979 ^{+0.0000000698} _{-0.0000000728}$} &  days \\
    \noalign{\smallskip}
    \quad Transit time & $T_0$ &$\mathcal{N}(2459446.49917,0.00019)$ & {\small $2459446.498519 ^{+0.000012} _{-0.000013}$} &  BJD\\
    \noalign{\smallskip}
    \quad Planet-to-star radius ratio & $R_p/R_\star$ &$\mathcal{U}(0,1)$ & {\small $0.15565 ^{+0.00024} _{-0.00021}$} &  -\\
    \noalign{\smallskip}
    \quad Transit depth & $\delta_t$ & - & {\small $24227 ^{+75} _{-66}$} &  ppm\\
    \noalign{\smallskip}
    \quad Impact parameter & $b$ &$\mathcal{U} (0,1)$ & {\small $0.6653 ^{+0.0021} _{-0.0020}$} &  - \\
    \noalign{\smallskip}
    \quad Scaled semi-major axis & $a/R_\star$ & - & {\small $8.8843 ^{+0.0173} _{-0.0177}$} &  -\\
    \noalign{\smallskip}
    \quad Occultation depth & $L$ &$\mathcal{U} (0,100)$ & $24.7 \pm 4.5$ &  ppm \\
    \noalign{\smallskip}
    \quad Geometric Albedo & $A_G$ & - & $0.076 \pm 0.016$ &  -\\
    \noalign{\smallskip}
    \quad Bond Albedo $3\sigma$-limits & $A_B$ & - & $0.013 \leq A_B \leq 0.42$ &  -\\
    \midrule
    Stellar Parameters &&&& \\
    \quad Stellar density & $\rho_\star$ &$\mathcal{N}(1.6246, 0.0936)$ & $1.9117 ^{+0.0112} _{-0.0114}$ &  $\rho_\odot$ \\
    \quad Limb darkening coefficients &&&& \\
    \multirow{2}{*}{\qquad CHEOPS passband} & $q_{1_{\mathrm{CHEOPS}}}$ &$\mathcal{N}(0.511, 0.05)$ & $0.439 ^{+0.026} _{-0.026}$ &  - \\
    \noalign{\smallskip}
    & $q_{2_{\mathrm{CHEOPS}}}$ & $\mathcal{N}(0.395, 0.05)$ &$0.387 ^{+0.037} _{-0.039}$ &  - \\
    \noalign{\smallskip}
    \multirow{2}{*}{\qquad TESS passband} & $q_{1_{\mathrm{TESS}}}$ & $\mathcal{N}(0.398, 0.05)$ & $0.358 ^{+0.019} _{-0.019}$ &  - \\
    \noalign{\smallskip}
    & $q_{2_{\mathrm{TESS}}}$ & $\mathcal{N}(0.340, 0.05)$ &$0.239 ^{+0.031} _{-0.029}$ &  - \\
    \bottomrule
    \end{tabular}
    \tablefoot{The Gaussian priors with mean $\mu$ and standard deviation $\sigma$ are displayed as $\mathcal{N}(\mu, \sigma)$. $\mathcal{U}(a,b)$ shows the uniform prior between $a$ and $b$.}
    \label{tab:fitted_transit_param}
\end{table*}

Before jointly analysing the transit observations of TESS and CHEOPS, we analysed these datasets individually. This was mainly to select and constrain the systematic and astrophysical noise models. In this analysis we did not assume any prior knowledge of planetary parameters except the period and the transit time, for which we used Gaussian priors based on values from \citet{Baluev2015}. We chose  this option because the \citet{Baluev2015} analysis uses a longer photometric baseline and radial velocity measurements to constrain these parameters, and therefore using these priors provides a more precise value than an analysis using only photometric observations by TESS and CHEOPS. We  also adopted a Gaussian prior on stellar density (see Section \ref{subsec_stellar} and Table \ref{table_stellar}). For the rest of the planetary parameters, we used wide uninformative priors (see  Table \ref{tab:fitted_transit_param}). To  parametrise the limb-darkening effect, we used the quadratic limb-darkening law with the efficient coefficient parametrisation proposed by \cite{2013MNRAS.435.2152K}.
We used the code of \citet{Espinoza2015} to compute Gaussian priors on the limb darkening coefficients (see Table \ref{tab:fitted_transit_param}). 
    
The individual TESS light curve was modelled using \texttt{juliet} \footnote{\url{https://github.com/nespinoza/juliet}} \citep{2019MNRAS.490.2262E}, which uses a transit model from \texttt{batman} \footnote{\url{https://github.com/lkreidberg/batman}} \citep{2015PASP..127.1161K}. In addition to the transit model, we added a jitter term, the mean out-of-transit flux, and a Gaussian process (GP) model to account for systematic and/or temporal astrophysical trends. The GP model was built using the Exponential-Mat\'{e}rn kernel in \texttt{juliet}, based on \texttt{celerite}\footnote{\url{https://github.com/dfm/celerite}} models \citep{Foreman-Mackey2017}. In practice, we first analysed only out-of-transit data with instrumental and GP models in order to better constrain the nuisance parameters and to obtain a better and fast convergence. We then used posteriors from this analysis as priors when we modelled the whole TESS dataset along with transit parameters.

\begin{figure}
    \centering
    \includegraphics[width=\hsize]{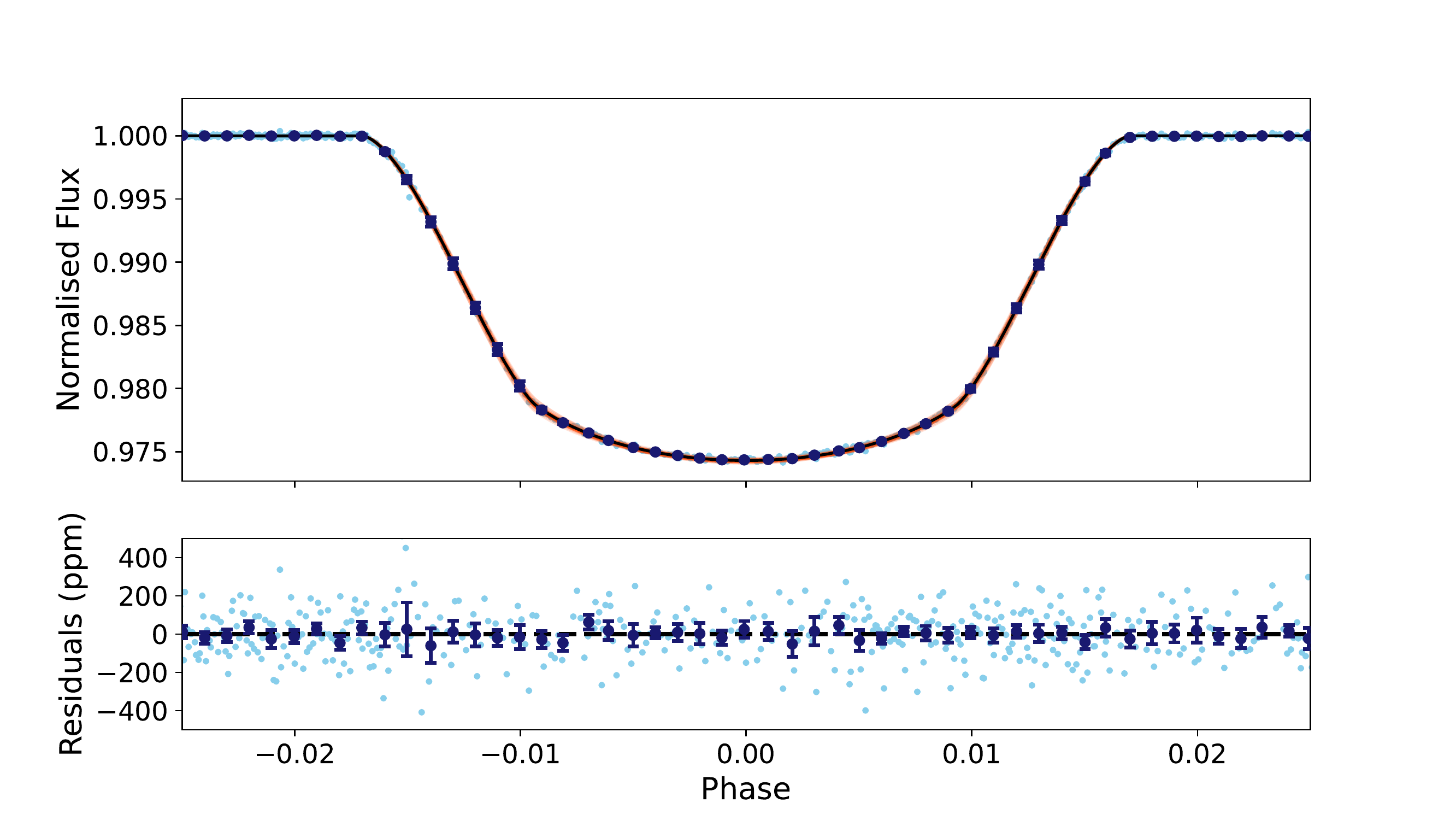}
    \includegraphics[width=\hsize]{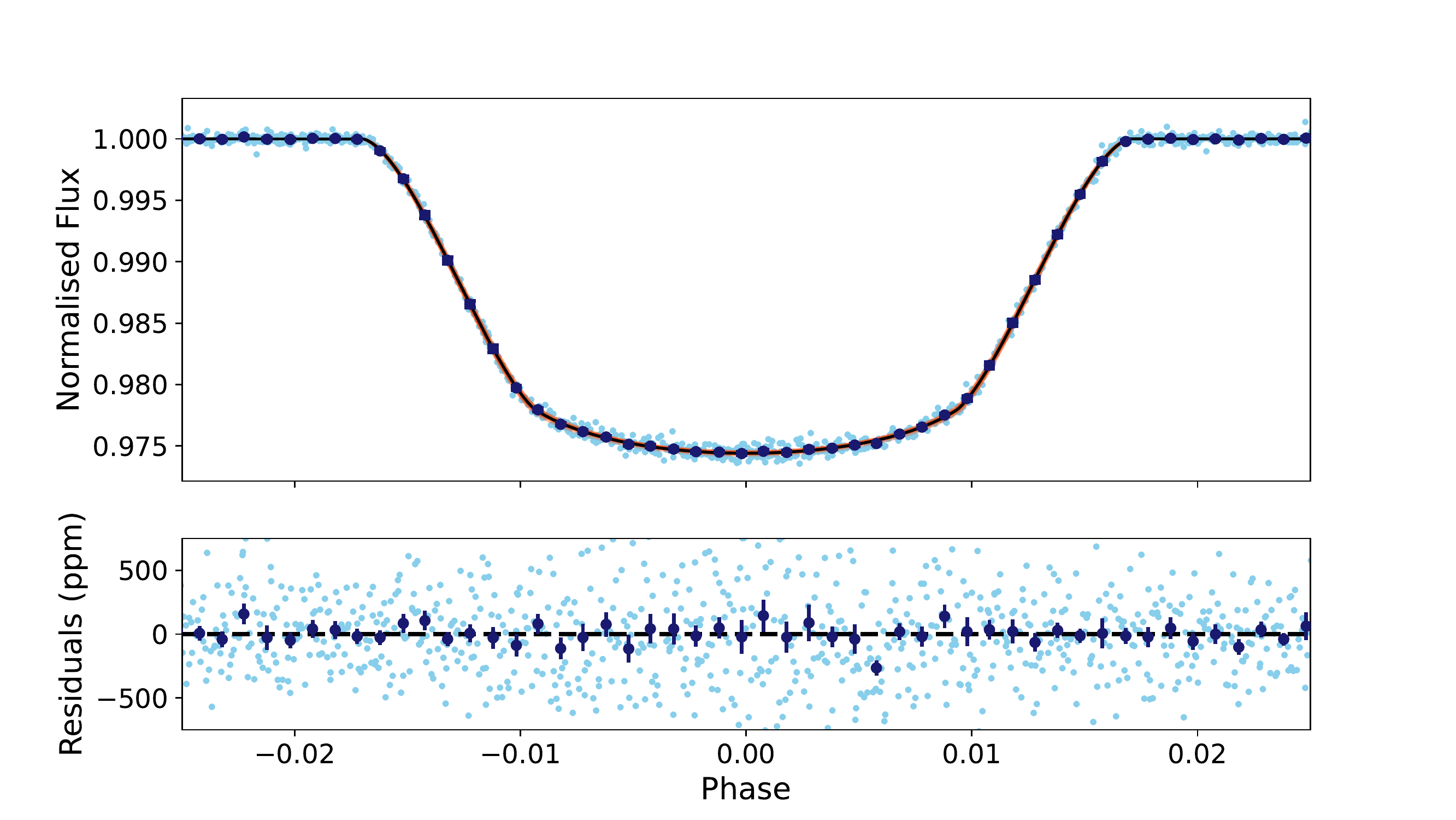}
    \caption{Top panel: Phase-folded and detrended transit light curve from two CHEOPS visits. The light and dark blue points are the original data and the 3-min binned data points, respectively. The black line is the median fitted model and the orange curves are the random models computed from the posterior samples. The residuals of the fit are shown below the light curve. Bottom panel: Same, but using TESS data observed during Sector 41.}
    \label{fig:transit}
\end{figure}
    
CHEOPS data are known to correlate with many instrumental parameters, mainly the roll angle, as the CHEOPS field of view rotates around the target during each of its orbits (see Section \ref{sec_data}). To correct for these trends we decorrelated the dataset against several instrumental parameters with the \texttt{PyCHEOPS}\footnote{\url{https://github.com/pmaxted/pycheops}} Python package, which was specifically developed to detrend and fit CHEOPS data \citep{Maxted2021}. We chose a set of detrending basis vectors to be included in the analysis by individually adding them one by one and retaining an additional vector only when supported by a higher Bayes factor computed from the Bayesian information criterion (BIC). This way we ended up using up to third-order harmonics of the roll angle and a second-order polynomial in the PSF centroid position. Both of the CHEOPS visits showed the well-known ramp effect caused by thermal effects in the instrument \citep{Maxted2021}. This effect manifests itself in the reduced data as an increased flux at the beginning of the observation, which quickly decreases to the same normalised flux level as the rest of the observation. We detrended this effect using the telescope-tube temperature as a correlated variable by adding a linear detrending model for the \texttt{thermfront2} parameter provided by the DRP to the detrending basis vectors. We then used \texttt{juliet} to actually fit the data with linear decorrelation against these parameters, and also a GP model to fit for the temporal trends. We followed the same two-step procedure to model the individual CHEOPS visits as we did with the TESS data. 

In the end, we modelled one sector of TESS data and two visits of CHEOPS jointly. In addition to the joint transit
model, the model that we used contained  a jitter term, an out-of-transit offset and a GP model for systematic and astrophysical trends for each dataset. The GP models were again built using the Exponential-Mat\'{e}rn kernel in \texttt{juliet}. We also included linear models to take care of instrumental correlations in two CHEOPS visits. The priors on these  noise parameters were Gaussian and were  based on our earlier analysis of individual datasets. We used nested sampling methods \citep{2004AIPC..735..395S, 10.1214/06-BA127, 2019S&C....29..891H}, as implemented in \texttt{juliet} via \texttt{dynesty}\footnote{\url{https://github.com/joshspeagle/dynesty}}  \citep{2020MNRAS.493.3132S} to sample from posteriors. The dataset and the median model, along with other randomly drawn models from posteriors, for CHEOPS and TESS are illustrated in Figure \ref{fig:transit}. The residuals show no significant long-term trend indicating the quality of our fit. The median posterior parameters, along with the 1$\sigma$ confidence interval, from our analysis are listed in Table \ref{tab:fitted_transit_param}.

We opted for   the \texttt{juliet} \texttt{Python} framework to fit all of the transit data as it allows for the simultaneous fitting of light curves from different instruments and the use of GPs with a variety of different kernels. Within the transit analysis \texttt{PyCHEOPS} was only used to determine the best detrending basis vectors for the CHEOPS data as it is especially designed to do so. 

\subsection{Occultation analysis}
\label{subsec_occultation_analysis}

\subsubsection{PyCHEOPS analysis}
We used the \texttt{PyCHEOPS} python package \citep{Maxted2021} to perform a combined analysis of all occultations. The package was chosen as it is optimised for fitting CHEOPS data. There was no need for the functionality of a multi-instrument analysis as the occultation dataset consists of only CHEOPS observations and we did not need the option of choosing between a variety of GP kernels as we did not use GPs in the occultation analysis (see Section \ref{subsubsec_discussion}). We defined Gaussian priors according to the results of the transit analysis (see Section \ref{subsec_transit_analysis}) on timing, period, depth, width, and impact parameter. For the stellar density we assumed the same prior as for the transit analysis (see Section \ref{subsec_stellar} and Table \ref{table_stellar}).

The data of each individual visit were corrected for the ramp effect (see Section \ref{subsec_transit_analysis}) with the \texttt{remove\textunderscore ramp} function built into \texttt{PyCHEOPS} \citep{Maxted2021}. We removed long-term time trends, with characteristic  timescales much larger than one CHEOPS orbit, with a second-order polynomial. For the polynomial fit we excluded the in-occultation data in order to not remove the planetary signal. Periodic flux changes correlated to the roll angle parameter (and consequently to the orbital position of the spacecraft) were treated with a two-step process.  First-order linear models of the sines and cosines of the roll angle $\theta$ and the angle $2 \theta$ were used to remove large-scale trends. Additionally, a 30-segment spline was fit on top of these models. The spline fits small-scale flux changes as a function of the roll angle to remove glint effects. We again exclude the in-occultation data from the spline fit in order to not remove the occultation signal. The glint-model is scaled by fitting an individual scale factor for each visit during the analysis. Analogously to the transit observations, we added additional linear models to detrend for smear noise, which is caused by electrons being smeared over several pixels during the read-out process of the CCD, x- and y-centroid offsets, which quantify the spacecraft jitter, contamination from background stars, and changes in background flux only if their addition was supported by a higher Bayes factor for the specific model.

Since we still observed flux changes correlated to time at characteristic timescales of about one CHEOPS orbit, we decided to apply additional time detrending. To do so we first used a subset of each individual visit containing only out-of-occultation data and ran  the MCMC method implemented in \texttt{PyCHEOPS} to fit for the parameters of the corresponding detrending model. The flux values of these subsets were then corrected for the instrumental systematics using the median values of the posterior distributions of these MCMC fits. Then we fitted  a spline to the corrected fluxes. The spline fitted to the out-of-occultation data of each individual visit was set up to have four segments; it used a fourth-order polynomial in each segment to account for the observed temporal flux-variations. Subsequently, the spline was added to the second-order polynomial that was applied to the dataset prior to fitting. Values of the spline during the occultation were interpolated from values before and after the occultation to ensure that  the planetary signal was not removed.

We decided to remove the sixth visit, performed on   20 July 2021, from our multi-visit analysis because as a result of Earth occultations this visit does not cover either ingress or egress. Both the second-order polynomial to remove long-term time trends and the spline fit to account for time trends at characteristic timescales of one CHEOPS orbit are fitted only on out-of-occultation data in order to not remove the planetary signal. Both functions are interpolated during the occultation event. Because this dataset is missing both ingress and egress, there are gaps before and after the in-occultation data. This results in a longer interpolation window than for other visits, making the fitted time trends less reliable. 

We then fitted each individual visit for the planetary parameters including the occultation depth, a white noise term, and the parameters of the detrending models using again the MCMC method implemented in \texttt{PyCHEOPS}. The fitted values of the detrending parameters of the individual visits were used as initial guesses of a multi-visit MCMC fit of all 12 remaining occultation observations. We fitted for a single occultation depth, the remaining planetary parameters, a white-noise term, and  individual detrending parameters for the detrending models and the glint scale for each visit. The parameters of the roll angle trend models were not explicitly calculated, but implicitly marginalised over in the MCMC following \citet{Rodrigo2017}. The multi-visit fit used a total of 53 free parameters. All fits also accounted for a light travel time delay of $31$ s \citep{Agol2010,Knutson2012}.

We measured an occultation depth in the CHEOPS bandpass ($330$ - $1100$ nm) of $24.7 \pm 4.5$ ppm. The close-up detrended, phase-folded, and fitted  light curve showing the occultation is plotted in Figure \ref{fig_lc_zoom}. The entire phase-folded light curve and the individual light curves can be found in the Appendix. To provide further evidence in favour of our result, we   repeated the whole procedure assuming a model without an occultation (i.e. an occultation depth equal to $0$). When comparing the two models we find that Baysian evidence supports the model with an occultation over the model without an occultation.

 \begin{figure}
   \centering
   \includegraphics[width=\hsize]{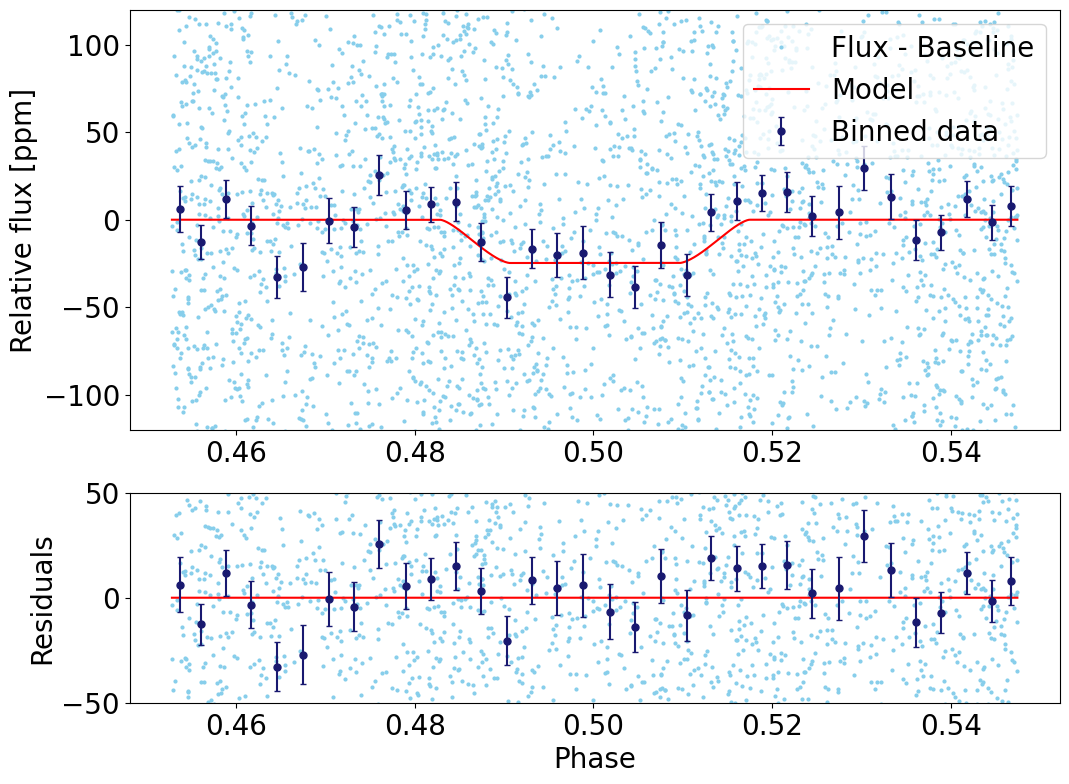}
      \caption{Close-up of the phase-folded, detrended, and fitted occultations of HD\,189733b observed by CHEOPS (top) and the residuals of the fitted model (bottom). The light blue points represent the individual data points, the dark blue points the  5 min bins, and the red line the model.}
         \label{fig_lc_zoom}
 \end{figure}
 
\subsubsection{Residual noise}
\label{subsubsec_discussion}
After the analysis of the acquired data, we observed some remaining residual noise: mainly short-term variations in stellar flux. These flux variations were particularly challenging to account for as their amplitudes and timescales were similar to that of the occultation signal, which has a duration of $108.6$ min. We attempted to fit GPs \citep[see e.g.][]{Gibson2012,Foreman-Mackey2017}, including the use of different kernels, on both the time and roll angle parameters as an alternative to the models previously used. We found that using a GP was not a suitable detrending approach for these specific observations because of the similarity of the amplitudes and timescales of the residual noise and the ingress--egress of the occultation. The GP model has no well-constrained GP hyperparameters, and we found that it tends to overfit the occultation events because it is not able to adequately distinguish possible stellar variations and the occultation signal. We decided not to remove flux variations with typical timescales smaller than one CHEOPS orbit and no significant correlation to any of the other instrumental parameters.

We further investigated if the residual noise could be due to the stellar granulation signals, as in \citet{Delrez2021} and \citet{flickerCHEOPS}. Unfortunately, our CHEOPS observations contain too many gaps and are too short (see Section \ref{subsec_occultation_data}) to properly identify and characterise this stellar noise source when studying periodograms.\footnote{According to Sulis et al. (2022) we would expect an increase in the power spectrum from high to low frequency with a cutoff frequency $f_g<600~\mu$Hz.} We then only compare the amplitude (RMS) of our residuals with the expected amplitude of this stellar noise. The RMS of the entire time series is about $95$ ppm, while if we bin the data over $4$ min (since the power spectrum on $<4 $ min timescales is dominated by photon noise), we get a typical RMS of around $32$ to $45$ ppm.
These amplitudes are similar to the amplitudes predicted by 3D hydrodynamic (HD) simulations of stellar granulation, which are $39.3$ ppm \citep{2022MNRAS.514.1741R}. However, without accessing the frequency constant of this residual noise, we cannot fairly conclude on a stellar origin. A white Gaussian noise of 95 ppm RMS could also lead to an RMS of around 40 ppm when binned over 4 min. The nature (instrumental or stellar origin) of this residual noise is therefore not well understood with the data at hand.


\section{Atmospheric properties of HD\,189733b}
\label{sec_atmos}

\subsection{Geometric albedo}
\label{sec_albedo}

The total occulted flux is composed of two major contributions: the thermal emission of the planet, which is determined by its equilibrium temperature, and the stellar light that is reflected by the planetary atmosphere. Therefore, to determine the reflective contribution the thermal emission has to be estimated. Following  \citet{Brandeker2022}, we   used the dayside temperature of 1192 $\pm$ 9 K, estimated with occultation measurements in the Spitzer 4.5 $\mu$m channel \citep{Knutson2012}, and extrapolate it to the CHEOPS bandpass. In the course of the extrapolation we  assumed an  irradiated atmosphere model provided by \citet{Molliere2015} with the parameters $T_{\mathrm{eff}}$ = 1250 $K$, [Fe/H] = 0.0, [C/O] = 0.70, planetary  $\log g$  = 3.0, and stellar spectral type K5. To model the stellar emission spectrum, we used a synthetic spectrum drawn from the \texttt{PHOENIX}\footnote{\url{https://phoenix.astro.physik.uni-goettingen.de}} library \citep{Husser2013}. The library contains synthetic spectra for a grid of stellar parameters. We chose the spectrum computed for a star with the parameters $T_{\mathrm{eff}}$ = 5000 $K$, [M/H] = 0.0 and stellar  $\log g$  = 4.5 as it most closely resembles HD\,189733 in the \texttt{PHOENIX} sample. The thermal contribution to the occultation depth is determined to be $1.42 \pm 0.03$ ppm.

The geometric albedo $A_g$ is then computed as \citep{Brandeker2022}

\begin{equation}
    A_g = \left( \frac{a}{R_p} \right)^2 \, L,
\end{equation}

\noindent with $a$ being the semi-major axis of the planet, $R_p$ the planetary radius, and $L$ the measured occultation depth corrected for the thermal contribution. In particular, $a/R_p$ was determined from the result of the transit fit (Table \ref{tab:fitted_transit_param}). The geometric albedo of HD\,189733b is calculated as

\begin{equation}
    A_g = 0.076 \pm 0.016
.\end{equation}

The measured value is consistent with the upper limit of $A_g<0.12$ measured in the 450 -- 570 nm Hubble Space Telescope, Space Telescope Imaging Spectrograph (HST STIS) band reported by \citet{Evans2013}. The achieved uncertainty of the measurement is identical to the uncertainty of the geometric albedo in the CHEOPS bandpass of HD\,209458b for which \citet{Brandeker2022} report $A_g = 0.096 \pm 0.016$.

\begin{figure}
   \centering
   \includegraphics[width=\hsize]{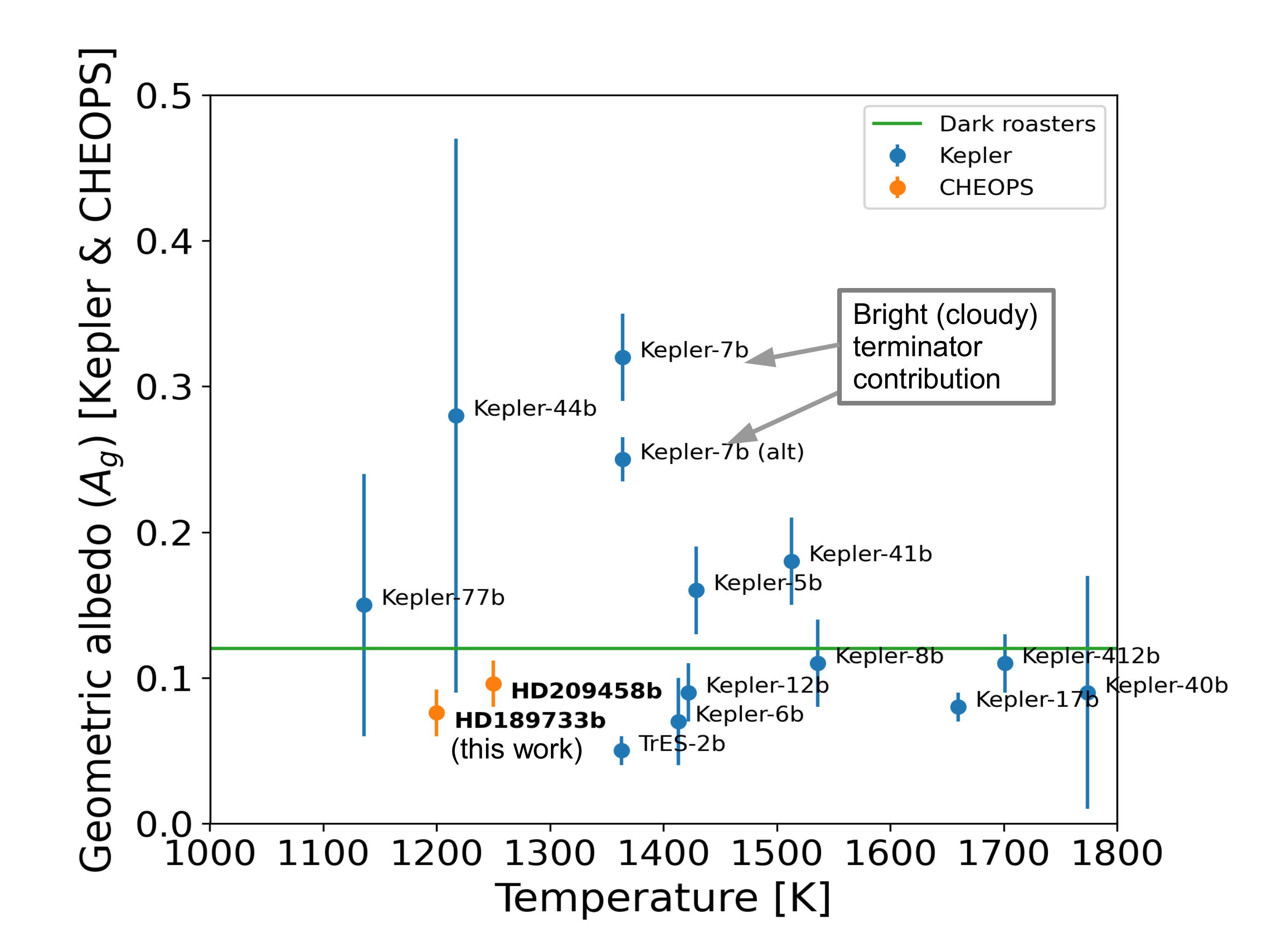}
      \caption{Geometric albedo of hot Jupiters measured by instruments sensitive in the mid-optical, i.e. Kepler (as reported by \citealt{Angerhausen2015}, blue) and CHEOPS (\citealt{Brandeker2022} and this work, orange) vs effective temperature. The green line at $A_{g}=0.12$ approximates the `dark roaster' regime with very low optical geometric albedos \citep{Sudarsky2000}. Kepler-7b is a prominent outlier,  even when taking into account  a recent update on its albedo (alt) \citep{Heng2021}. Its unusually bright albedo may be explained by reflection of clouds at the planetary limbs \citep{Adams2022}.}
         \label{fig: Ag_Kepler_Cheops}
 \end{figure}

\subsection{Atmospheric Na content}
The measured low geometric albedo of HD\,189733b in the optical wavelength range makes this planet part of the hot Jupiter population ($T_{\mathrm{eq}} =$ 1000 -- 1800 K) for which low geometric albedos were measured in the Kepler and CHEOPS optical wavelength range (Figure \ref{fig: Ag_Kepler_Cheops}). This would make HD\,189733b like HD\,209458b a typical dark and cloud-free `Class IV roaster' \citep{Sudarsky2000,Seager2000},  a planet for which the low albedo in the optical is consistent with a model that assumes that atmospheres are cloud-free at the pressure level where Rayleigh scattering occurs on the dayside\footnote{Rayleigh scattering in hot Jupiters occurs typically at $\approx$ 10~mbar \citep{Vidal-Madjar2011}.} and also assumes for the absorption in the CHEOPS (and  in the Kepler) band to be dominated by the resonant Na doublet. 

Following the same approach as in \citet{Brandeker2022} taken from \citet{Heng2021}, we computed the theoretical geometric albedo of HD\,189733b as a function of atmospheric Na content using the above model. We assumed a dayside temperature of 1200~K consistent with Spitzer secondary eclipse observations by \citet{Knutson2007}, \citet{Agol2010}, and \citet{Todorov2014}. Furthermore, a cloud-free hydrogen dominated atmosphere was assumed, where we considered $\mathrm{H_2O}$ and Na as dominant absorbers in the CHEOPS bandpass as outlined in \citet{Brandeker2022}. Using elemental abundances for the host star as listed in Table~\ref{table_stellar}, we derived for $\mathrm{H_2O}$ a volume mixing ratio of $5.4 \times 10^{-4}$ in chemical equilibrium using \citet{Heng2016}.

\begin{figure}
   \centering
   \includegraphics[width=\hsize]{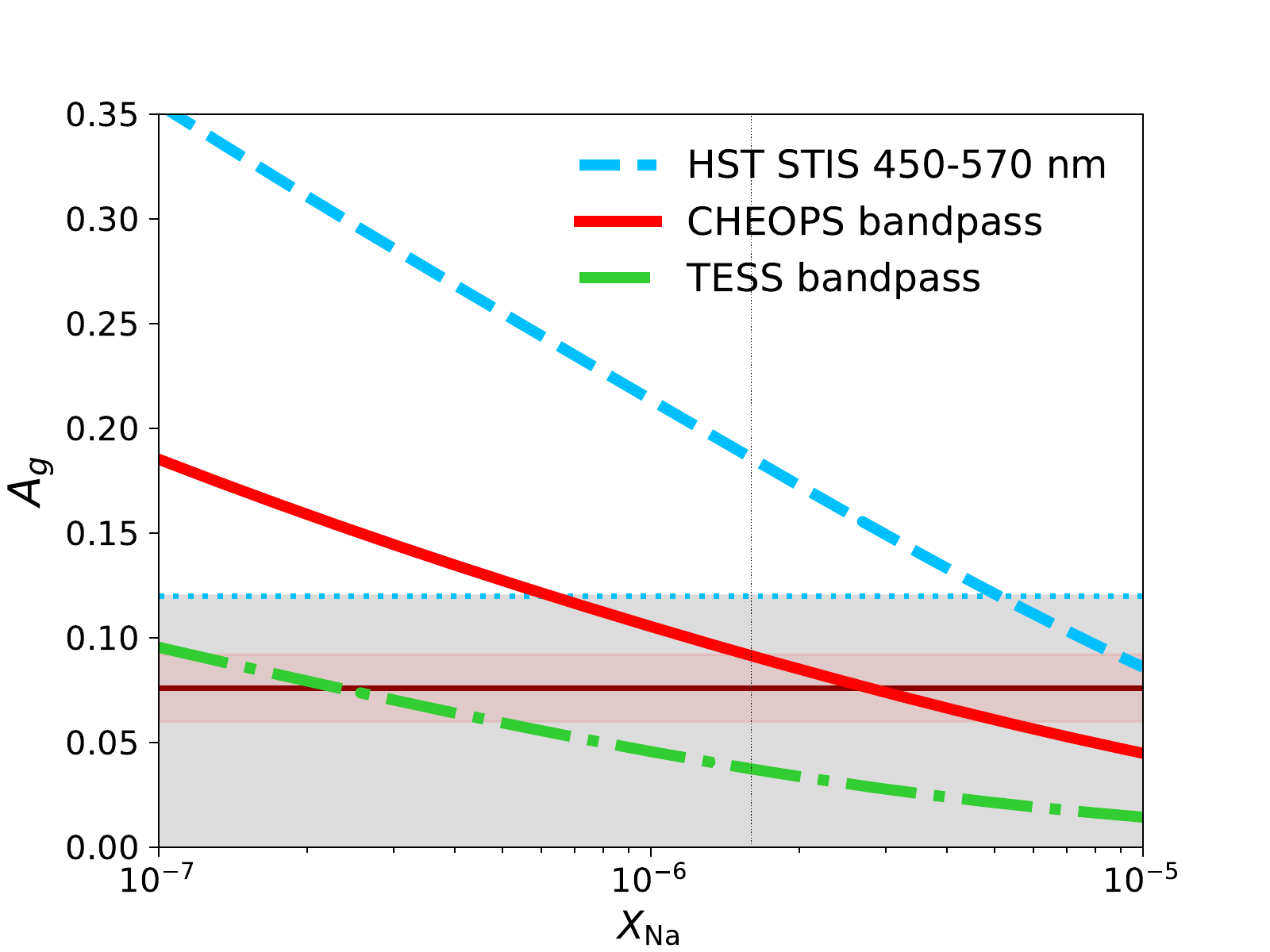}
      \caption{Comparison of measured and theoretical geometric albedos of HD\,189733 b. The blue dotted line denotes the upper limit for the geometric albedo $A_g=0.12$ as measured by \citet{Evans2013} in the HST STIS bandpass. The red shaded region denotes our CHEOPS albedo measurement ($\pm 1 \sigma$). The blue dashed, red, and green curves respectively denote for HST STIS, CHEOPS, and TESS the theoretically calculated albedos $A_g$ integrated over the respective bandpasses for different $X_{\mathrm{Na}}$ abundances. HST STIS and CHEOPS measurements are consistent with each other if $X_{\mathrm{Na}} \geq 6\times 10^{-6}$. The corresponding TESS albedo would be very low, and thus undetectable according to the model used here.}
         \label{fig: Comparing albedos}
 \end{figure}
 
Comparing the theoretical $A_g$ calculations for different $X_{\mathrm{Na}}$ contents integrated over the CHEOPS, HST STIS, and TESS bandpass (Figure~\ref{fig: Comparing albedos}) shows that our CHEOPS albedo measurement is consistent with sodium elemental abundances between $1-3\times$ its stellar value. When combining the measured CHEOPS and HST STIS albedos \citep{Evans2013}, a super-stellar sodium elemental abundance ($X\approx 3$) is required for the observations to be consistent with the calculated model values. The corresponding albedo in the TESS bandpass would be very low, and thus probably not detectable.

\subsection{Bond albedo}
\label{sec_bond}
Occultation observations at optical wavelengths can also be used to infer lower and upper boundaries of the Bond albedo \citep{Rowe2006}. They can therefore complement Bond albedo measurements obtained from thermal phase curves. \citet{Schwartz2015} show that by assuming either zero or perfect reflection for all wavelengths outside of the observed bandpass, lower and upper limits for the Bond albedo can be derived from the spherical albedo ($A_S$) in the observed bandpass. The spherical albedo also covers all directions, but in contrast to the Bond albedo, it is defined at specific wavelengths. Following \citet{Schwartz2015}, the Bond albedo can be computed as

\begin{equation}
\label{eq:Ab}
    A_B = A_S^{\mathrm{ opt}} \cdot f_{\star}^{\mathrm{ opt}} + A_S^{\mathrm{oob}} \cdot(1-f_{\star}^{\mathrm{ opt}}),
\end{equation}

\noindent where $A_S^{\mathrm{ opt}}$ is the spherical albedo at optical wavelengths, $f_{\star}^{\mathrm{ opt}}$ is the ratio of stellar flux emitted at optical wavelengths   to the total emitted flux, and $A_S^{\mathrm{oob}}$ is the spherical albedo for all wavelengths outside of the observed bandpass. To derive lower and upper limits on the Bond albedo, $A_S^{\mathrm{oob}}$ is assumed to be 0 and 1, respectively. In the case of the CHEOPS bandpass, this results in

\begin{equation}
    A_B^{\mathrm{min}} = A_S^{\mathrm{CH}} \cdot f_{\star}^{\mathrm{CH}},
\end{equation}

\begin{equation}
    A_B^{\mathrm{max}} = A_S^{\mathrm{CH}} \cdot f_{\star}^{\mathrm{CH}} + 1\cdot(1-f_{\star}^{\mathrm{CH}}),
\end{equation}

\noindent where $f_{\star}^{\mathrm{CH}}$ is the ratio of stellar flux emitted in the CHEOPS bandpass   to the total emitted flux, and $A_S^{\mathrm{CH}}$ is the spherical albedo of the planet in the CHEOPS bandpass. Using the same \texttt{PHOENIX} spectrum as in Section \ref{sec_albedo} to model the stellar emission, we find $f_{\star}^{\mathrm{CH}} = 0.637$ for HD\,189733.  The geometric and spherical albedo are linked to each other by

\begin{equation}
    \label{eq_q}
    A_S = q A_G
\end{equation}

\noindent with $q$ depending on the exact reflective qualities of the atmosphere \citep{Pollack1986,Burrows2010}. \citet{Heng2021} derive $q = 0.761$ for Rayleigh single scattering and isotropic multiple scattering under the assumption of the single scattering albedo to be unity. As the low geometric albedo measured in the CHEOPS bandpass implies Rayleigh scattering to be dominant at the height in the atmosphere where optical light is scattered (see Section \ref{sec_albedo}), we apply $q = 0.761$ to transfer our measured geometric albedo in the CHEOPS bandpass to a spherical albedo:

\begin{equation}
    A_S^{\mathrm{CH}} = 0.058 \pm 0.012
.\end{equation}Accounting for $3\sigma$ confidence intervals of our geometric albedo measurement from CHEOPS, this results in

\begin{equation}
\begin{split}
    A_B^{\mathrm{min}} & = 0.013,\\
    A_B^{\mathrm{max}} & = 0.42.
\end{split}
\end{equation}

All previous measurements of the Bond albedo of HD\,189733b using thermal phase curves are consistent with the lower and upper boundaries derived from the CHEOPS geometric albedo. Using the same approach, \citet{Schwartz2015} report a slightly higher value for the  lower limit of $A_B^{\mathrm{min}} = 0.043$ from secondary eclipse observations in the Hubble 290 -- 450 nm channel. However, this value does not take into account confidence intervals, and assumes $q=5/4$ and a black-body spectrum for the stellar emission. In comparison, when assuming our median value for $A_g^{CH}$ and $A_S^{\mathrm{oob}} = 0$, we retrieve $A_B^{\mathrm{min}} = 0.037$. Although the measured geometric albedo derived from short-wavelength Hubble data was high \citep{Evans2013}, those data add little information when constraining the Bond albedo. The bandpasses mostly overlap with only the wavelength range of 290 -- 350 nm being covered by Hubble alone. The integrated stellar emission flux at these wavelengths amounts to only about 1.4\% of the total stellar emission and is therefore negligible. We note that the \texttt{PHOENIX} model predicts less emission at very short optical and long UV wavelengths compared to a black-body spectrum.

\citet{Schwartz2015} also suggest assuming $A_S^{\mathrm{oob}} = 0.5$ for hot Jupiters, which would result in $A_B^{0.5} = 0.22 \pm 0.08$. On the other hand, it is also possible to take Bond albedos inferred from thermal phase curves and calculate the corresponding $A_S^{\mathrm{oob}}$ values. Since the stellar fluxes at wavelengths shorter than the CHEOPS band are mostly negligible for HD\,189733, $A_S^{\mathrm{oob}}$ can also be used as a proxy of the spherical albedo at infrared wavelengths. Assuming $A_B = 0.37 \pm 0.05$ from \citet{Schwartz2015}  results in a peculiarly high out-of-band albedo of $A_S^{\mathrm{oob}} = 0.92 \pm 0.16$. Using $A_B = 0.16 \pm 0.11$ from \citet{Keating2019} results in a more moderate out-of-band albedo of $A_S^{\mathrm{oob}} = 0.34 \pm 0.32$. 

\begin{figure}
   \centering
   \includegraphics[width=\hsize]{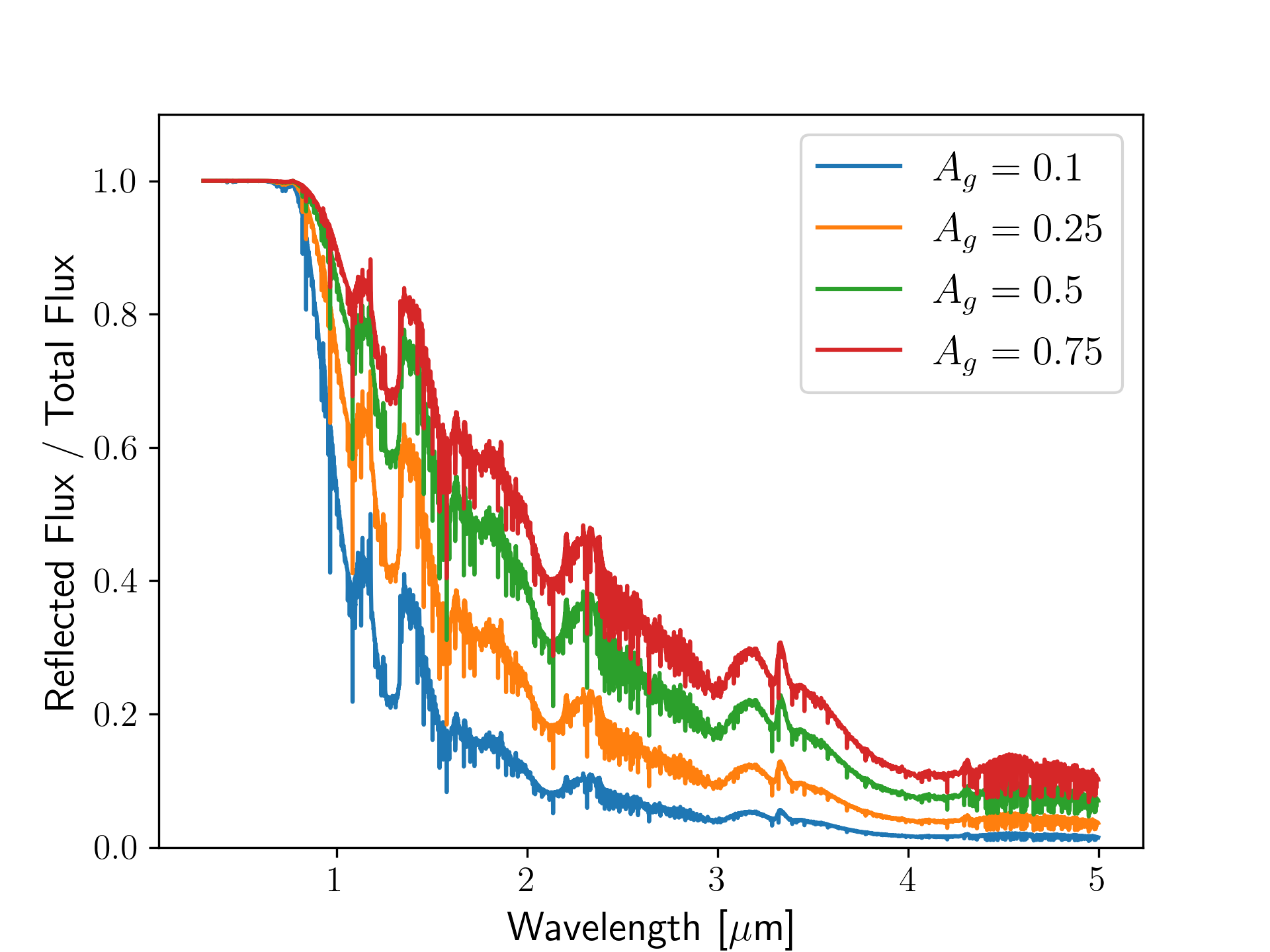}
      \caption{Contribution of reflected light to the total planetary flux for different grey geometric albedos as a function of wavelength. For the stellar emission spectrum a \texttt{PHOENIX} model \citep{Husser2013} with the parameters $T_{\mathrm{eff}}$ = 5000 $K$, [M/H] = 0.0, and  $\log g$  = 4.5 was used. To model the planetary atmosphere an irradiated atmosphere model provided by \citet{Molliere2015} with the parameters $T_{\mathrm{eff}}$ = 1250 $K$, [Fe/H] = 0.0, [C/O] = 0.70, planetary  $\log g$  = 3.0, and stellar spectral type K5 was adopted. The thermal emission from the planet was extrapolated from the dayside temperature of 1192 $\pm$ 9 K, estimated with occultation measurements in the Spitzer 4.5 $\mu$m channel \citep{Knutson2012}.}
         \label{fig:thermal_albedos}
 \end{figure}

Since the measured light from a planet is composed of both light emitted by the planet and reflected stellar light, observations of thermal phase curves have to account for both reflected and emitted light. Figure \ref{fig:thermal_albedos} shows the estimated contribution of reflected light to the total planetary flux for different grey geometric albedos as a function of wavelength for the HD\,189733 system for a given planetary dayside temperature. Here we used the same \texttt{PHOENIX} and irradiated atmosphere models as in Section \ref{sec_albedo} to model the stellar emission and planetary atmosphere respectively. Analogously to determining the thermal emission of the planet in the CHEOPS bandpass, we again extrapolated thermal emission at a given wavelength from the dayside temperature of 1192 $\pm$ 9 K, estimated with occultation measurements in the Spitzer 4.5 $\mu$m channel \citep{Knutson2012}. Contrary to the usual assumption that the vast majority of infrared flux is contributed by thermal emission from the planet, in the case of a high geometric albedo the reflective component at longer wavelengths remains significant. If this contribution to the occultation depth is neglected or underestimated when interpreting infrared observations, this leads to overestimated thermal flux, and thus to overestimated temperatures. Earlier models that retrieve higher Bond albedo estimates at values of roughly $0.4$, as reported by \citet{Schwartz2015} and \citet{Schwartz2017}, imply peculiarly high reflectivity at infrared wavelengths. In this case, reflected light needs to be properly accounted for in the phase curve analysis, which itself becomes more complex, as the absolute amount of reflected light is larger, and therefore small inaccuracies in its treatment lead to more significant deviations from the true value. On the other hand, models resulting in more moderate estimates, like the measurement $A_B = 0.16_{-0.10}^{+0.11}$ by \citet{Keating2019}, seem to suffer much less from this problem.

\section{Conclusion}
\label{sec_conclusion}
We reported the measurement of the geometric albedo of the hot Jupiter HD\,189733b in the CHEOPS bandpass ($A_g = 0.076 \pm 0.016$) by analysing 13 observations of its occultation. We started by refining the transit parameters using both TESS and CHEOPS observations  (see Table \ref{tab:fitted_transit_param}). Subsequently, we fitted for the occultation depth ($L = 24.7 \pm 4.5$ ppm) and estimated the contribution of thermal emission from the planet to the total occulted flux. Finally, we inferred the geometric albedo in the CHEOPS bandpass. The measured value is consistent with previous measurements of the geometric albedo of this target and other gas giants, which are often found to be $\lesssim 0.1$ \citep{Angerhausen2015,Esteves2015,Heng2013}. The achieved precision of the measurement further proves the capability of CHEOPS to detect low-amplitude occultation signals, just as was done for HD\,209458b \citep{Brandeker2022}. 

We interpreted the measurement with atmospheric models assuming a cloud-free atmosphere at the pressure level where Rayleigh scattering occurs on the dayside, as well as varying Na enhancement of the planet compared to its host star. We compared these models to the CHEOPS detection as well as the HST STIS upper limit \citep{Evans2013}. Both measurements are consistent with a super-stellar Na elemental abundance ($X\approx 3$). In this case the geometric albedo in the TESS band would be very low, making a secondary eclipse of HD\,189733b in TESS data unlikely to be detectable. Although consistent within $1\sigma$ confidence intervals, CHEOPS measurements hint at HD\,189733b having a slightly lower geometric albedo in the CHEOPS band than HD\,209548b \citep{Brandeker2022}. This is matched, in the framework of the applied models, by a higher required Na abundance as the planetary brightness at optical wavelengths is keenly sensitive to absorption in the broad Na doublet. Finally, following \citet{Schwartz2015}, we present lower and upper limits at $3\sigma$ confidence on the Bond albedo of the planet at 0.013 and 0.42, respectively. When compared with Bond albedos derived from thermal phase curve observations, we note that earlier   higher estimates of the Bond albedo would also require peculiarly high reflectance at infrared wavelengths, while the more recent lower estimates lead to more consistent values.

\begin{acknowledgements}
We thank the referee, N. B. Cowan, for his valuable comments and suggestions.
CHEOPS is an ESA mission in partnership with Switzerland with important contributions to the payload and the ground segment from Austria, Belgium, France, Germany, Hungary, Italy, Portugal, Spain, Sweden, and the United Kingdom. The CHEOPS Consortium would like to gratefully acknowledge the support received by all the agencies, offices, universities, and industries involved. Their flexibility and willingness to explore new approaches were essential to the success of this mission.
This paper includes data collected with the TESS mission, obtained from the MAST data archive at the Space Telescope Science Insti- tute (STScI). Funding for the TESS mission is provided by the NASA Ex- plorer Program. STScI is operated by the Association of Universities for Re- search in Astronomy, Inc., under NASA contract NAS 5-26555.
This work has been carried out within the framework of the National Centre of Competence in Research PlanetS supported by the Swiss National Science Foundation under grants 51NF40\textunderscore182901 and 51NF40\textunderscore205606. The authors acknowledge the financial support of the SNSF.
ML acknowledges support of the Swiss National Science Foundation under grant number PCEFP2\_194576.
This project was supported by the CNES.
L.C. acknowledges funding from the European Union H2020-MSCA-ITN-2019 under Grant Agreement no. 860470 (CHAMELEON).
SS acknowledges financial support from the Programme National de Planétologie (PNP) and the Programme National de Physique Stellaire (PNPS) of CNRS-INSU.
ABr was supported by the SNSA.
This work was supported by FCT - Fundação para a Ciência e a Tecnologia through national funds and by FEDER through COMPETE2020 - Programa Operacional Competitividade e Internacionalizacão by these grants: UID/FIS/04434/2019, UIDB/04434/2020, UIDP/04434/2020, PTDC/FIS-AST/32113/2017 \& POCI-01-0145-FEDER- 032113, PTDC/FIS-AST/28953/2017 \& POCI-01-0145-FEDER-028953, PTDC/FIS-AST/28987/2017 \& POCI-01-0145-FEDER-028987, O.D.S.D. is supported in the form of work contract (DL 57/2016/CP1364/CT0004) funded by national funds through FCT.
S.G.S. acknowledge support from FCT through FCT contract nr. CEECIND/00826/2018 and POPH/FSE (EC).
YA and MJH acknowledge the support of the Swiss National Fund under grant 200020\_172746.
We acknowledge support from the Spanish Ministry of Science and Innovation and the European Regional Development Fund through grants ESP2016-80435-C2-1-R, ESP2016-80435-C2-2-R, PGC2018-098153-B-C33, PGC2018-098153-B-C31, ESP2017-87676-C5-1-R, MDM-2017-0737 Unidad de Excelencia Maria de Maeztu-Centro de Astrobiologí­a (INTA-CSIC), as well as the support of the Generalitat de Catalunya/CERCA programme. The MOC activities have been supported by the ESA contract No. 4000124370.
S.C.C.B. acknowledges support from FCT through FCT contracts nr. IF/01312/2014/CP1215/CT0004.
XB, SC, DG, MF and JL acknowledge their role as ESA-appointed CHEOPS science team members.
ACC acknowledges support from STFC consolidated grant numbers ST/R000824/1 and ST/V000861/1, and UKSA grant number ST/R003203/1.
The Belgian participation to CHEOPS has been supported by the Belgian Federal Science Policy Office (BELSPO) in the framework of the PRODEX Program, and by the University of Liège through an ARC grant for Concerted Research Actions financed by the Wallonia-Brussels Federation; L.D. is an F.R.S.-FNRS Postdoctoral Researcher.
B.-O.D. acknowledges support from the Swiss National Science Foundation (PP00P2-190080).
This project has received funding from the European Research Council (ERC) under the European Union’s Horizon 2020 research and innovation programme (project {\sc Four Aces}; grant agreement No 724427). AD and DE acknowledges financial support from the Swiss National Science Foundation for project 200021\_200726.
MF and CMP gratefully acknowledge the support of the Swedish National Space Agency (DNR 65/19, 174/18).
DG gratefully acknowledges financial support from the CRT foundation under Grant No. 2018.2323 ``Gaseousor rocky? Unveiling the nature of small worlds''.
M.G. is an F.R.S.-FNRS Senior Research Associate.
SH gratefully acknowledges CNES funding through the grant 837319.
KGI is the ESA CHEOPS Project Scientist and is responsible for the ESA CHEOPS Guest Observers Programme. She does not participate in, or contribute to, the definition of the Guaranteed Time Programme of the CHEOPS mission through which observations described in this paper have been taken, nor to any aspect of target selection for the programme.
This work was granted access to the HPC resources of MesoPSL financed by the Region Ile de France and the project Equip@Meso (reference ANR-10-EQPX-29-01) of the programme Investissements d'Avenir supervised by the Agence Nationale pour la Recherche.
PM acknowledges support from STFC research grant number ST/M001040/1.
LBo, GBr, VNa, IPa, GPi, RRa, GSc, VSi, and TZi acknowledge support from CHEOPS ASI-INAF agreement n. 2019-29-HH.0.
This work was also partially supported by a grant from the Simons Foundation (PI Queloz, grant number 327127).
IRI acknowledges support from the Spanish Ministry of Science and Innovation and the European Regional Development Fund through grant PGC2018-098153-B- C33, as well as the support of the Generalitat de Catalunya/CERCA programme.
GyMSz acknowledges the support of the Hungarian National Research, Development and Innovation Office (NKFIH) grant K-125015, a a PRODEX Experiment Agreement No. 4000137122, the Lend\"ulet LP2018-7/2021 grant of the Hungarian Academy of Science and the support of the city of Szombathely.
V.V.G. is an F.R.S-FNRS Research Associate.
NAW acknowledges UKSA grant ST/R004838/1.
This research has made use of the Spanish Virtual Observatory (https://svo.cab.inta-csic.es) project funded by MCIN/AEI/10.13039/501100011033/ through grant PID2020-112949GB-I00.
\end{acknowledgements}

\bibliographystyle{aa} 
\bibliography{AA202245016.bib} 

\begin{appendix}
\onecolumn
\section{Supplementary material}

\begin{figure}[H]
   \centering
   \includegraphics[width=\hsize]{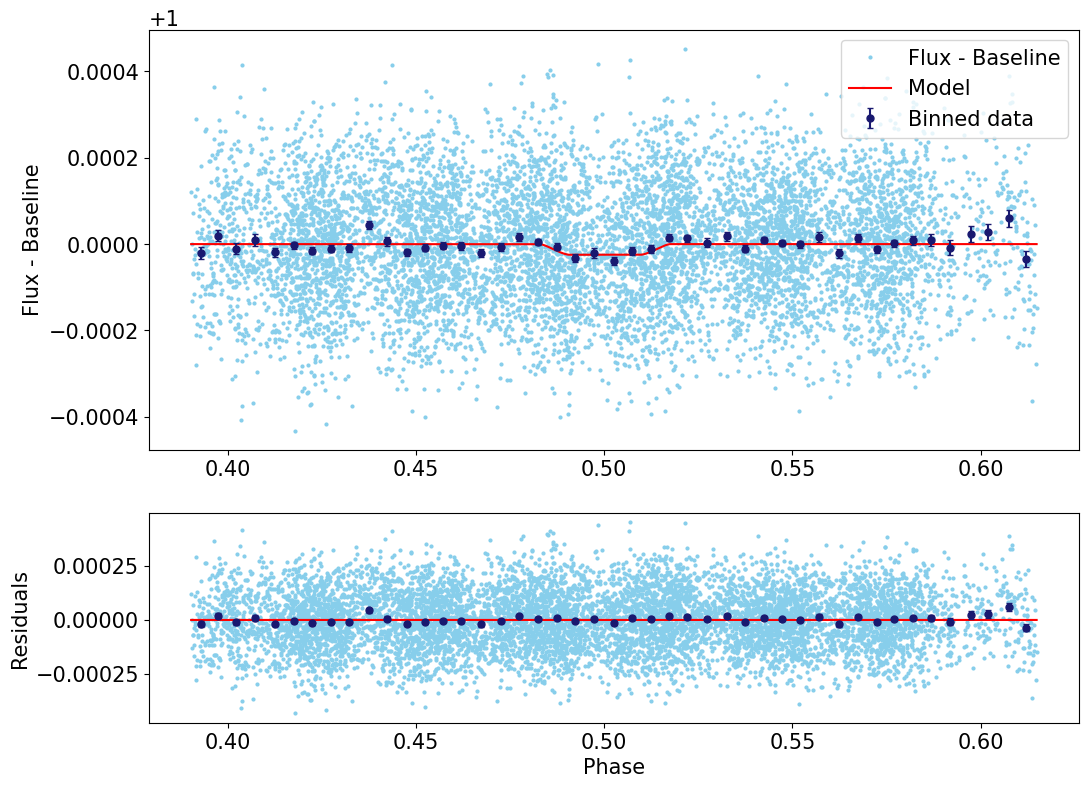}
   \caption{Phase-folded, detrended, and fitted light curve of occultation observations of HD\,189733b performed by CHEOPS. The dark blue points represent the 7 min binned data.}
              \label{fig_lc}%
\end{figure}

   \begin{figure}
   \centering
   \includegraphics[width=0.5\hsize]{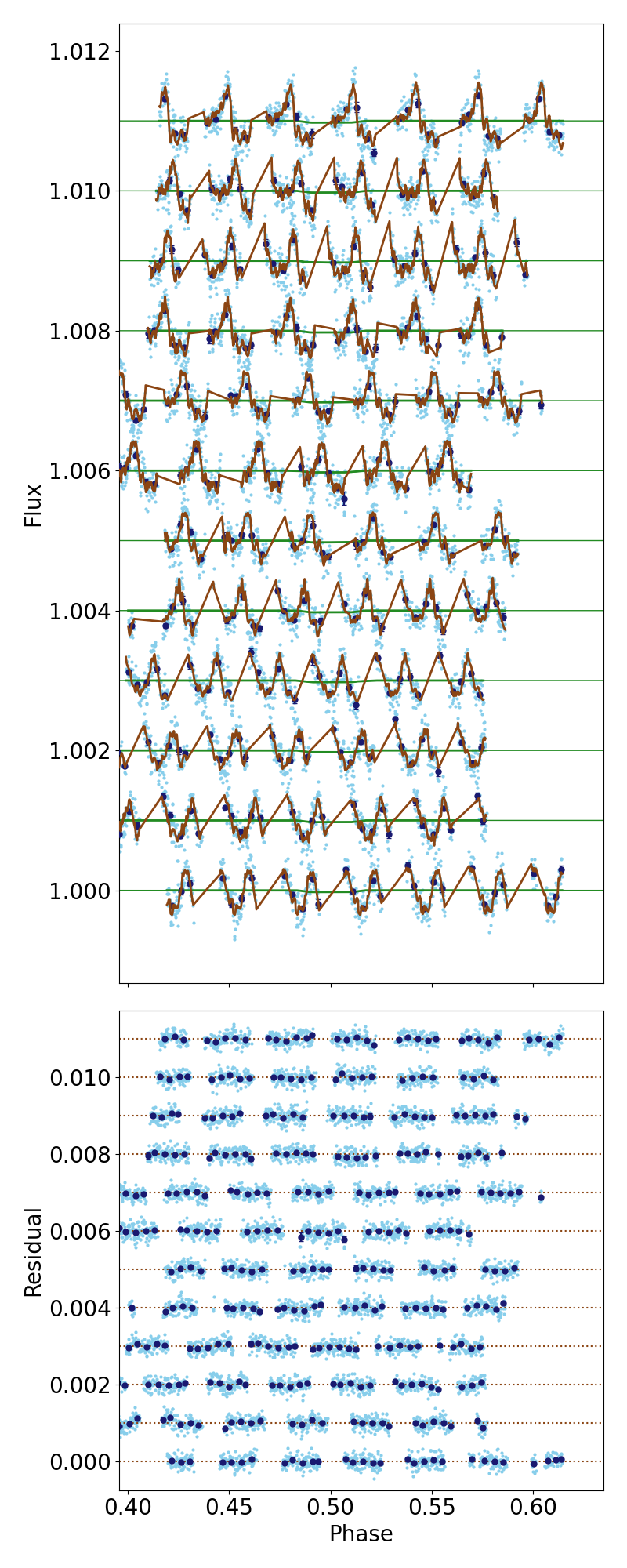}
   \caption{Light curves of the occultation observations of HD\,189733b performed by CHEOPS. The brown line is the fitted model. The displayed fluxes are already corrected for time trends with characteristic timescales equal to or larger than one CHEOPS orbit (roughly 100 minutes), which were removed prior to fitting.}
              \label{fig_mcmc}%
    \end{figure}

   \begin{figure}
   \centering
   \includegraphics[width=0.5\hsize]{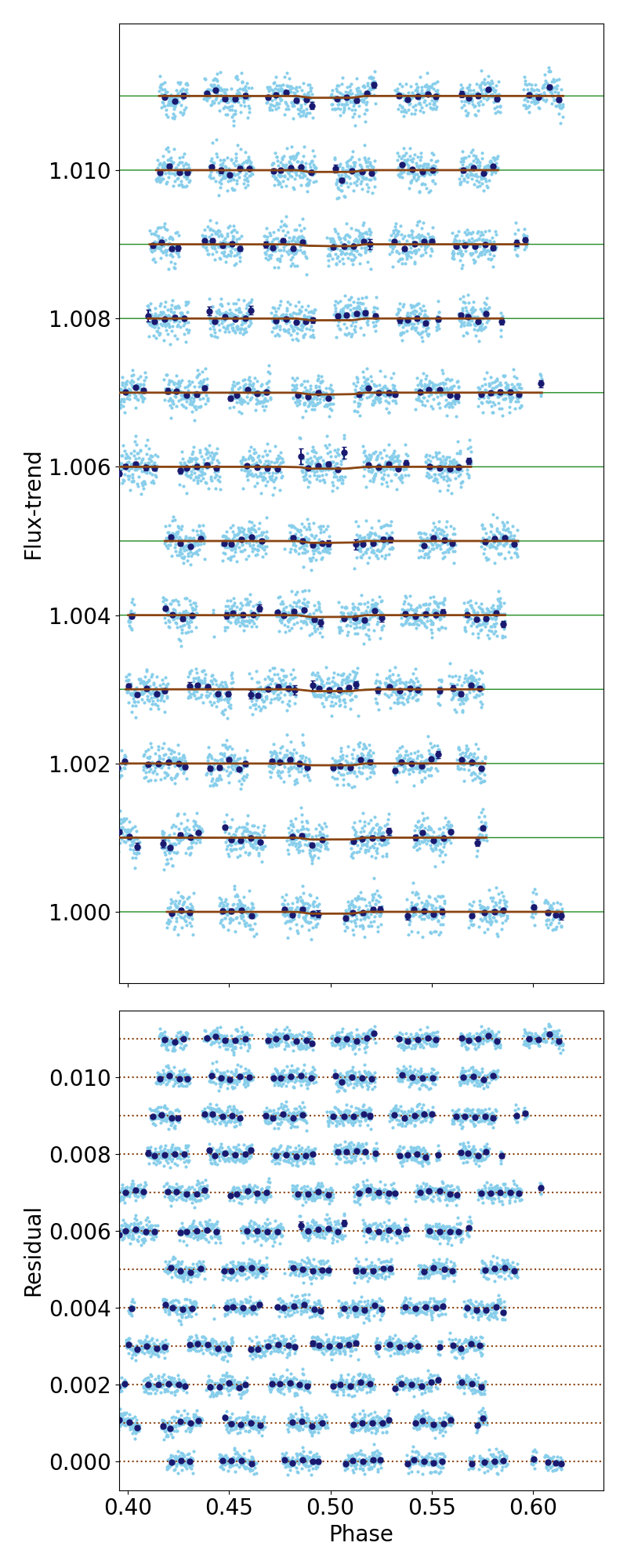}
   \caption{Detrended and fitted light curves of the occultation observations of HD\,189733b performed by CHEOPS.}
              \label{fig_mcmc_detrended}%
    \end{figure}

\begin{figure}
    \centering
    \includegraphics[width=\hsize]{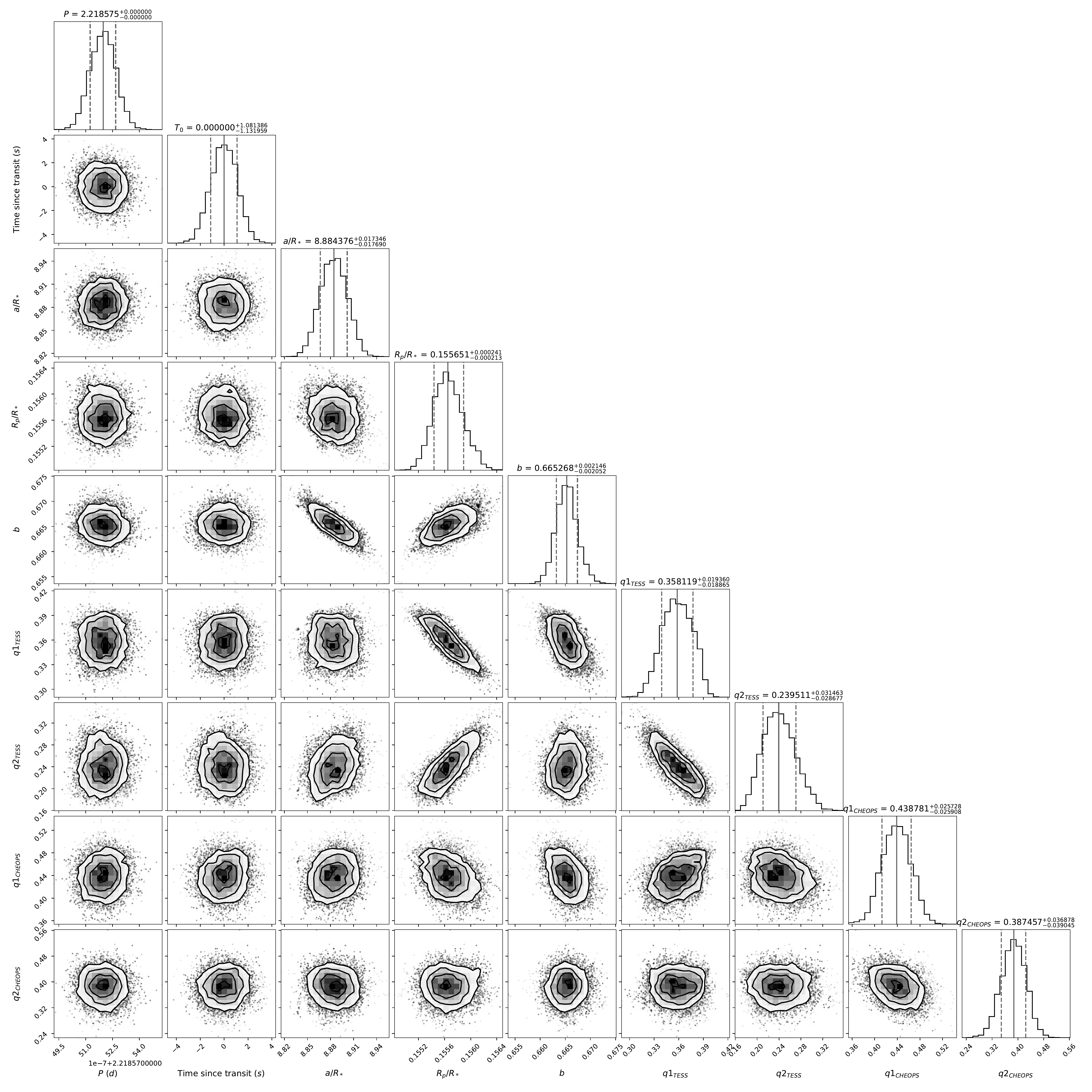}
    \caption{Corner plot for the fitted parameters from our transit analysis.}
    \label{fig:corner_transit}
\end{figure}
    
\end{appendix}

\end{document}